\documentclass[11pt]{article}

\usepackage[top=1in,bottom=1in,left=1in,right=1in]{geometry}

\usepackage{graphicx}

\usepackage{times}

\title{Human collective intelligence as distributed Bayesian inference}

\author
{Peter M. Krafft$^{1\ast}$, Julia Zheng$^{2}$, Wei Pan$^{2}$, Nicol\'{a}s Della Penna$^{3}$,\\ Yaniv Altshuler$^{2}$, Erez Shmueli$^{2,4}$, Joshua B. Tenenbaum$^{1,5}$, Alex Pentland$^{2}$\\
\\
\normalsize{$^{1}$Computer Science and Artificial Intelligence Laboratory, MIT, Cambridge, MA, USA}\\
\normalsize{$^{2}$MIT Media Lab, Cambridge, MA, USA}\\
\normalsize{$^{3}$Research School of Computer Science, Australian National University, Canberra, Australia}\\
\normalsize{$^{4}$Department of Industrial Engineering, Tel-Aviv University, Tel-Aviv, Israel}\\
\normalsize{$^{5}$Department of Brain and Cognitive Sciences, MIT, Cambridge, MA, USA}\\
\\
\normalsize{$^\ast$To whom correspondence should be addressed; e-mail:  pkrafft@mit.edu.}
}

\date{}

\begin{document}

\maketitle

  \begin{abstract}
    Collective intelligence is believed to underly the remarkable
    success of human society. The formation of accurate shared beliefs
    is one of the key components of human collective intelligence. How
    are accurate shared beliefs formed in groups of fallible
    individuals? Answering this question requires a multiscale
    analysis. We must understand both the individual
    decision mechanisms people use, and the properties and dynamics of
    those mechanisms in the aggregate. As of yet, mathematical tools
    for such an approach have been lacking.  To address this gap, we
    introduce a new analytical framework: We propose that groups
    arrive at accurate shared beliefs via distributed Bayesian
    inference.  Distributed inference occurs through information
    processing at the individual level, and yields rational belief
    formation at the group level.  We instantiate this framework in a
    new model of human social decision-making, which we validate using
    a dataset we collected of over 50,000 users of an online social
    trading platform where investors mimic each others' trades using
    real money in foreign exchange and other asset markets. We find
    that in this setting people use a decision mechanism in which
    popularity is treated as a prior distribution for which decisions
    are best to make. This mechanism is boundedly rational at the
    individual level, but we prove that in the aggregate implements a
    type of approximate ``Thompson sampling''---a well-known and
    highly effective single-agent Bayesian machine learning algorithm
    for sequential decision-making.  The perspective of distributed
    Bayesian inference therefore reveals how collective rationality
    emerges from the boundedly rational decision mechanisms people use.
    %% Hence, while the decision mechanisms that
    %% people use may be boundedly rational at the individual level, they
    %% can still lead to collectively rational group behavior.
    %% Our analysis exemplifies a new class of
    %% computational models of human collective behavior in which
    %% boundedly rational decision mechanisms provably lead to
    %% collectively rational group-level belief formation.
    %no individual alone is responsible for the complete computation performed by the group.
    %% We find that these individuals implement in aggregate a type of Thompson smapling
    %% This dataset
    %% provides a unique window into a large community's collective
    %% learning process since it includes both the social information and
    %% the objective evidence decision-makers have.
    %% The framework we
    %% introduce could be applied widely across complex social domains
    %% beyond this specific instance, and could also lead to new computer
    %% algorithms for distributed Bayesian inference.
\end{abstract}

Human groups have an incredible capacity for technological,
scientific, and cultural creativity.  Our historical accomplishments
and the opportunity of modern networked society to stimulate ever
larger-scale collaboration have spurred broad interest in
understanding the problem-solving abilities of groups---their
collective intelligence.  The phenomenon of collective intelligence
has now been studied extensively across animal species
\cite{couzin2009collective}; collective intelligence has been argued
to exist as a phenomenon distinct from individual intelligence in
small human groups \cite{woolley2010evidence}; and the remarkable
abilities of large human collectives have been extensively documented
\cite{surowiecki2005wisdom}.  However, while the work in this area has
catalogued what groups can do, and in some cases the mechanisms behind
how they do it, we still lack a coherent formal perspective on what
human collective intelligence actually is.  There is a growing view of
group behavior as implementing distributed algorithms
\cite{hutchins1995cognition,kearns2006experimental,couzin2007collective,feinerman2013theoretical}, which goes a
step beyond the predominant analytical framework of agent-based models
in that it formalizes specific information processing tasks that
groups are solving.  Yet this perspective provides little
insight into one of the key features of human group cognition---the
formation of shared beliefs
\cite{mesmer2009information,cattuto2009collective,theiner2010recognizing,rendell_why_2010,mason_collaborative_2012,rzhetsky2015choosing}.
%% How are the shared beliefs of groups formed, any why are
%% they at times adaptive, and at times maladaptive?

  %In contrast to previous models that have only captured
  %isolated group-level properties such as convergence or evolutionary
  %fitness guarantees in restricted settings,

%% Social learning---the ability for people
%% to share beliefs through learning from others---is central to
%% individual and collective problem-solving
%% \cite{rendell_why_2010,mason_collaborative_2012}, and hence an
%% adequate analytical framework for understanding group cogntion must
%% provide a formal account of how social learning relates to collective
%% intelligence.

\begin{figure*}
  \centering
  \includegraphics[width = 0.45\linewidth]{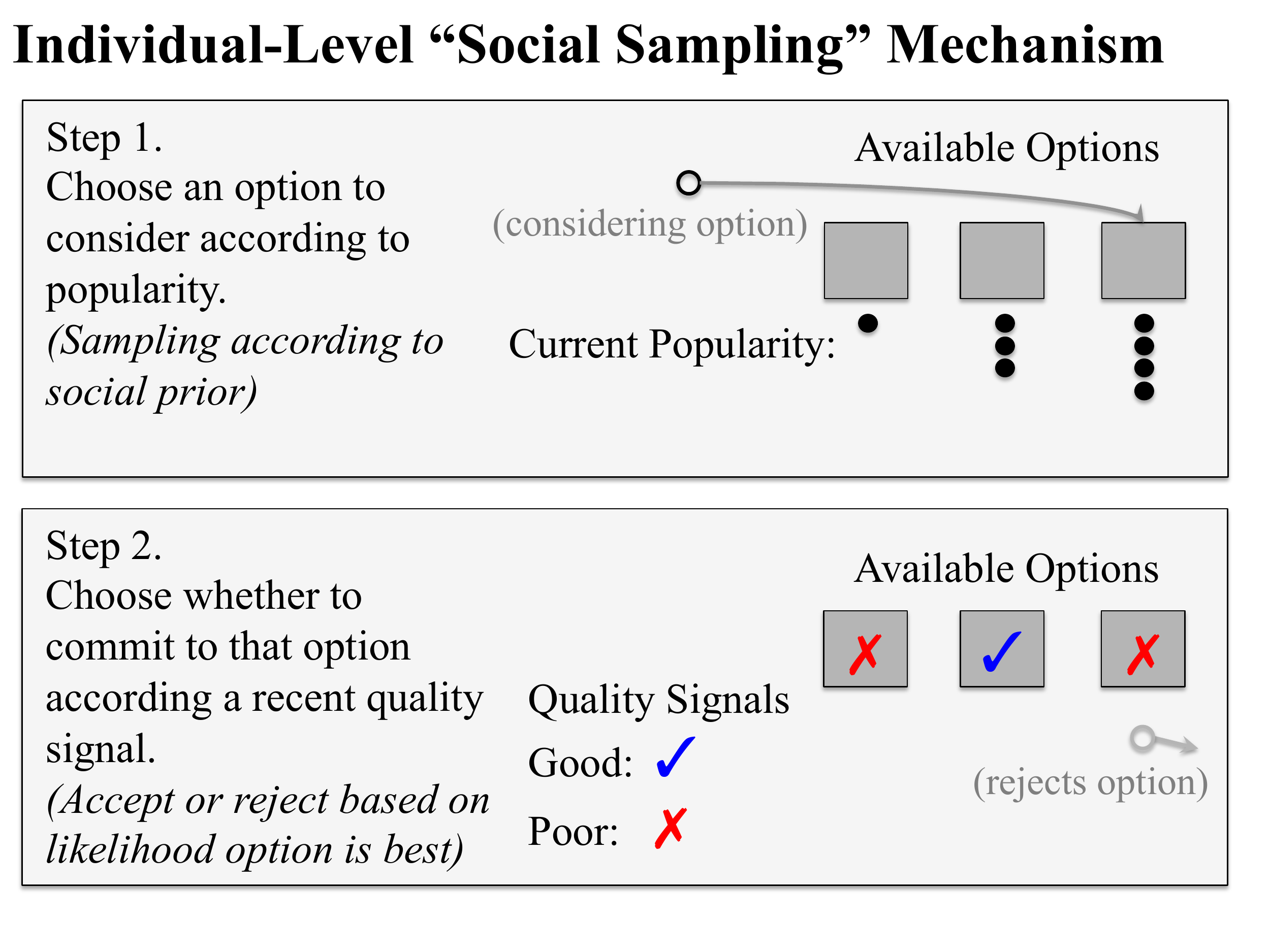}
  \includegraphics[width = 0.45\linewidth]{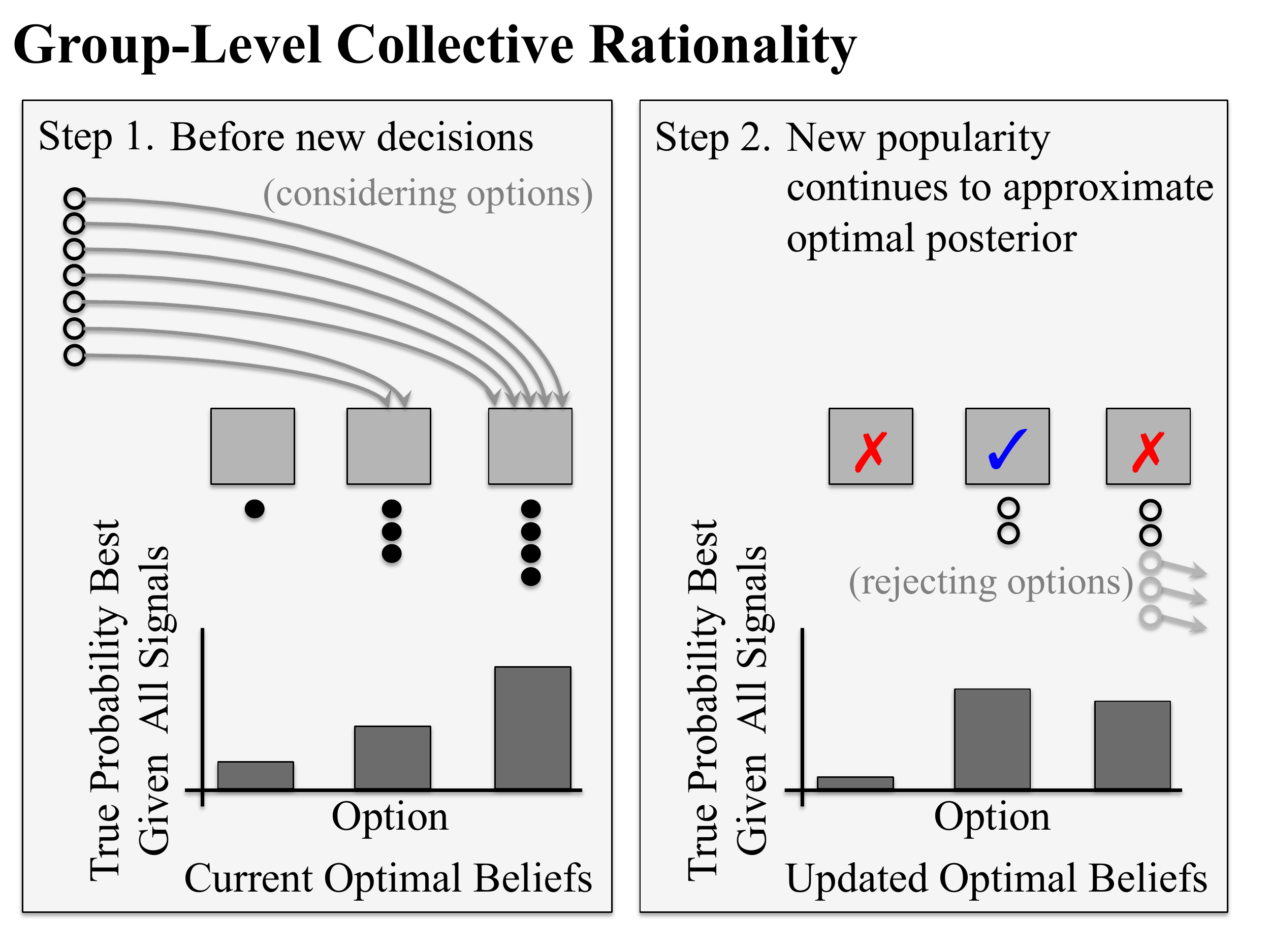}
  \caption{A schematic illustration of the ``social sampling''
    mechanism we propose as a model of human social decision-making,
    and an illustration of how this mechanism yields collectively
    rational belief formation at the group level.  Individuals treat
    current popularity as a prior distribution, and sample according
    to this prior in order to choose an option to consider taking.  An
    individual then commits to a considered option with probability
    proportional to the likelihood that the option is best given a
    recent objective signal of quality.  If current popularity
    approximates the current optimal posterior distribution that each
    option is best given all previous quality signals, then when a
    large group of decision-makers continues to use the social
    sampling mechanism, popularity will continue to approximate the
    optimal posterior.}
  \label{fig:model}
\end{figure*}

Recent work in cognitive science has provided a way to understand
belief formation at the level of individual intelligence.  One
productive framework treats people as approximately Bayesian agents
with rich mental models of the world
\cite{griffiths2006optimal,griffiths_bayesian_2008}.  Beliefs that
individuals hold are viewed as posterior distributions, or samples
from these distributions, and the content and structure of those
beliefs come from the structure of people's mental models as well as
their objective observations.  Belief formation is viewed as
approximate Bayesian updating, conditioning these mental models on an
individual's observations.  We propose to model human collective
intelligence as distributed Bayesian inference, and we present the
first empirical evidence for such a model from a large behavioral
dataset.  Our model shows how shared beliefs of groups can be formed
at the individual level through interactions with others and private
boundedly rational Bayesian updating, while in aggregate implementing
a rational Bayesian inference procedure.

We instantiate this broader framework in the context of social
decision-making.  A social decision-making problem represents a
setting in which a group of people makes decisions among a shared set
of options, and previous decisions are public.  Examples of social
decision-making problems include choosing what restaurant to visit
using a social recommendation system, choosing how to invest your
money after reading the news, or choosing what political candidate to
support after talking with your friends.
%going to restaurants, or yelp restaurant reviews or music downloads,
%or investment decisions, or endorsing political candidates, or betting
%on who will win the super bowl, things like that.
We instantiate our framework of collective intelligence as distributed
Bayesian inference in a new mathematical model of human social
decision-making, and then we evaluate this model quantitatively with a
unique large-scale social decision-making dataset that allows us to
test the model's predictions in ways previous datasets have not
enabled.

%We propose a two-stage model of social decision-making that affords a
%coherent cognitive interpretation at the group level.

Our model, illustrated in Figure \ref{fig:model}, posits that
individuals first choose options to consider based on popularity, then
choose to commit to those options based on assessments of objective
evidence.  A Bayesian interpretation of this strategy suggests that
people are trying to infer the best decisions to make by treating
popularity as a prior distribution that summarizes past evidence.  A
large group of people using this strategy will collectively perform
rational inferences.  We test this model, and thereby test our broader
framework, by showing the model is able to account for the patterns of
social influence we observe in our data better than several
alternatives.
%% We then explore further implications of this
%% model through simulations.

Formally, we propose that a person in a social decision-making context
will first choose an option $j$ to consider with probability
proportional to its current popularity at time $t$, $p_{j,t}$.  The
decision-maker then evaluates the quality of that option using a
recent performance signal $r_{j,t}$, which for simplicity we assume
indicates either good quality ($r_{j,t} = 1$) or poor quality
($r_{j,t} = 0$).  The decision-maker then chooses to commit to that
option with probability $\eta > 0.5$ if the signal is good or $1 -
\eta$ if the signal is bad (where $\eta$ is a free parameter).  The
probability that a decision-maker commits to option $j$ at time $t$ is
then
\begin{equation}
\frac{p_{j,t} \eta^{r_{j,t}} (1 - \eta)^{1 - r_{j,t}}}{\sum_k p_{k,t} \eta^{r_{k,t}} (1 - \eta)^{1 - r_{k,t}}}.
\end{equation}

We call this decision-making strategy ``social sampling''.  Heuristics
similar to the social sampling model have been previously proposed
\cite{bianconi_bose_2001,krumme_quantifying_2012,pentland2014social,
  sumpter_quorum_2009,mcelreath2008beyond,beheim_strategic_2014,granovskiy2015integration}.
In particular, two-stage social decision-making models appear to be
common across animal species
\cite{pratt_agent-based_2005,seeley_group_1999}.  The social sampling
model is also mathematically equivalent to a novel kind of stochastic
actor-oriented model \cite{snijders1996stochastic}, and is related to
a number of other prior models of social learning and social influence
\cite{degroot1974reaching,bikhchandani_theory_1992,golub2010naive,castellano2009statistical,granovetter1978threshold}.
However, our treatment is substantially different from these previous
accounts.  The specific mathematical form of this model, which in its
details is distinct from any previously proposed, has a boundedly
rational cognitive interpretation at the individual level, and affords
a coherent cognitive interpretation at the group level.
%also
%allows us to analyze collective rationality at the group level.
We can thus establish a formal relationship between individual
behavior, individual cognition, and expressed collective belief.

From the perspective of an individual decision-maker, there is a
simple Bayesian cognitive analysis \cite{griffiths_bayesian_2008} that
explains the mathematical form of Equation (1).  Equation (1) can be
interpreted as the posterior probability that option $j$ is the best
option available if the mental model people have is that (a) there is
a single best option (a needle in the haystack) producing good signals
with probability $\eta$ while all other options produce good and bad
signals with equal probability, and (b) the market share of option
$j$, $\frac{p_j}{\sum_k p_k}$, corresponds to the prior probability
that $j$ is the best option.  (See Section ``Social Sampling Model
Specification'' in the appendix for additional details.)  Assuming
that popularity is a reasonable proxy for prior probability, choosing
option $j$ with the probability given by Equation (1),
i.e. ``probability matching'' on this posterior, can be viewed as
rational under certain resource constraints
\cite{vul2014one,gershman2015computational}.  Social sampling is
therefore a boundedly rational probabilistic decision-making strategy
that is far more computationally efficient than exhaustive search over
all options.

Taking into consideration the behavior of an entire group that uses
social sampling, we can also prove that it is rational for
decision-makers to treat popularity as a prior distribution in this
way.  When the entirety of a population uses this strategy, the
expected new popularity of each option $p_{j,t'}$ will be proportional
to the previous posterior probability that the option was best.
Popularity will then come to approximate a rational prior in the
steady-state dynamics of a large population all participating in
social sampling.  Social sampling becomes ``collectively rational''
for this reason.  Optimal posterior inferences arise from a calibrated
social prior represented at the group level being combined with new
evidence obtained by the group members.  The group therefore
collectively transforms a boundedly rational heuristic at the
individual level into a distributed implementation of a fully Bayesian
inference procedure at the group level.  In light of the Bayesian
perspective of cognition, the collective rationality of social
sampling shows how groups can be interpreted as irreducible,
cognitively coherent, distributed information processing systems.

\section*{Results}

To test the social sampling model, we make use of a unique
observational dataset from a large online financial trading platform
called ``eToro''.  eToro's platform allows users to invest real money
in foreign exchange markets and other asset markets, and also allows
its users to mimic each other's trades.  When a trader becomes a
``mimicker'' of a target user, the mimicking trader allocates a fixed
amount of funds to automatically mirror the trades that the target
user makes.  The number of mimickers a user has represents the
popularity of the decision to mimic that user, and the user's trading
performance indicates the quality of the decision to mimic that user.
We treat the decision of whom to mimic on eToro as a social
decision-making problem.
The dataset we use consists of a year of trading activity from 57,455
users of eToro's online social trading platform
\cite{pan_decoding_2012}.  The website includes search and profile
functionality that display both popularity and performance information
of the users on the site.  For our analysis, we summarize each user's
trading performance with expected return on investment (ROI) from
closed trades over a 30-day rolling window, which is similar to the
performance metrics presented to the site's users.

\begin{figure*}
  \centering
  \includegraphics[width = 0.45\linewidth,trim={0 3in 0.5in 3in}, clip]{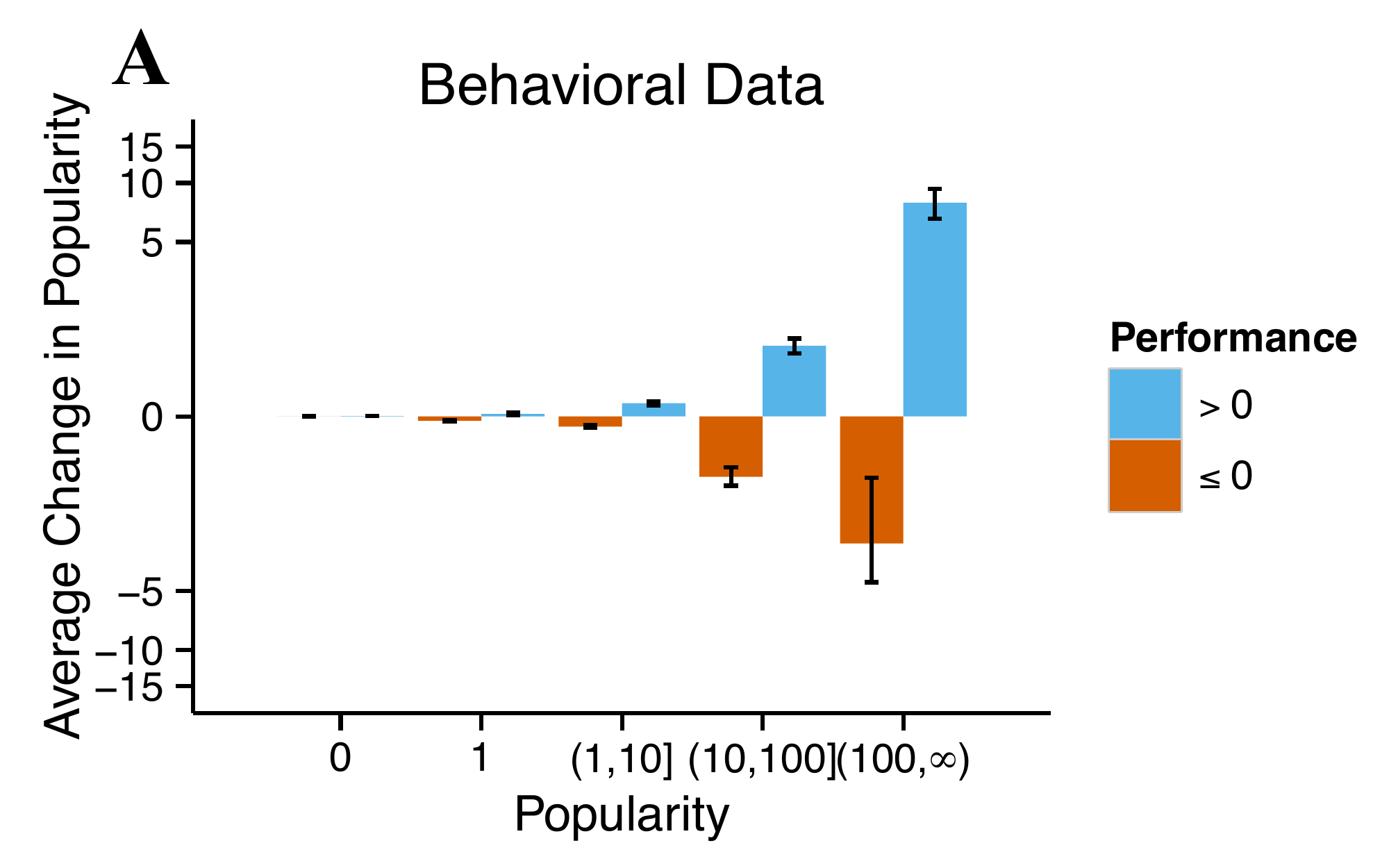}
  \\
  \includegraphics[width = 0.325\linewidth,trim={0 3in 2.5in 3in}, clip]{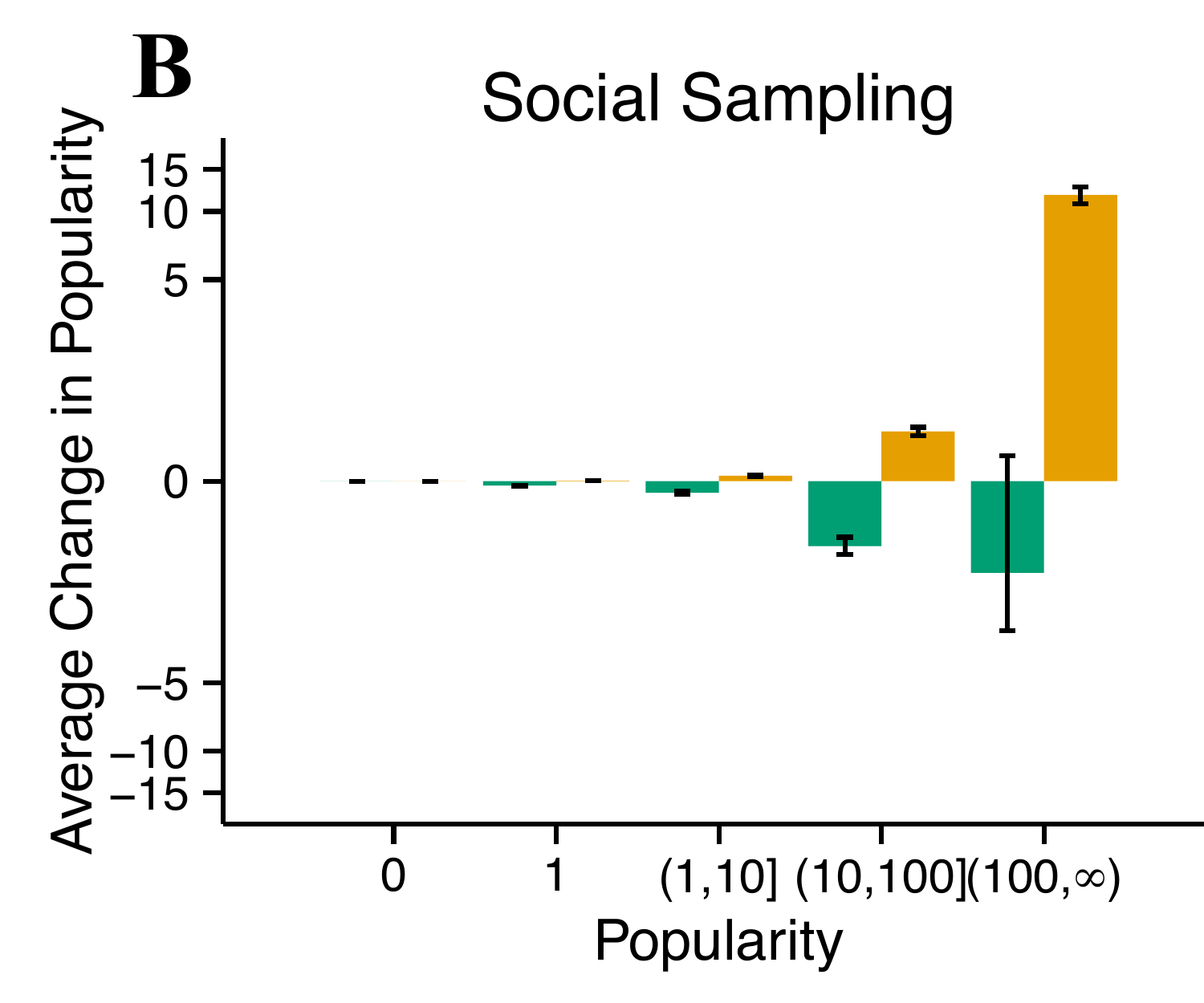}
  \includegraphics[width = 0.325\linewidth,trim={0 3in 2.5in 3in}, clip]{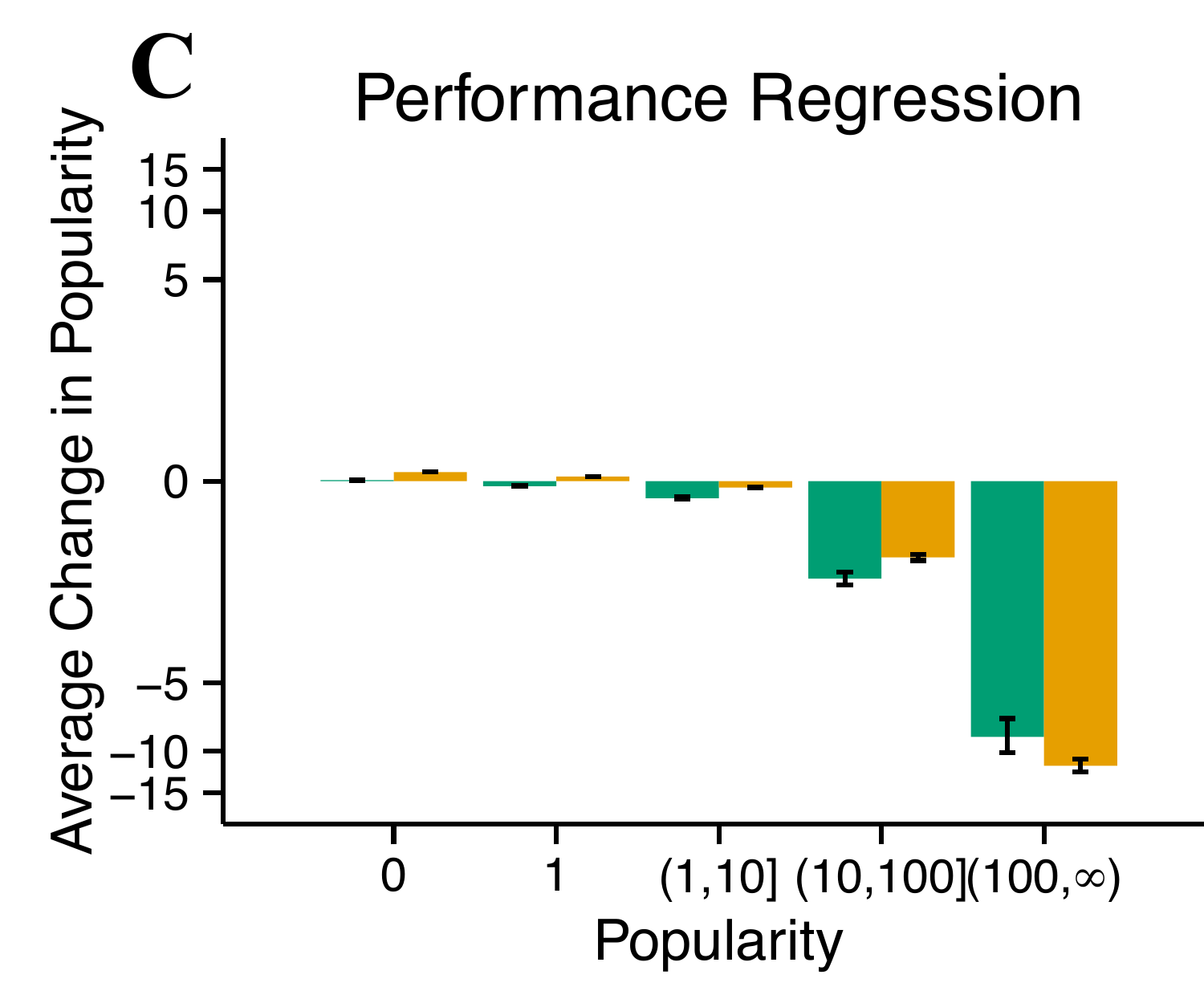}
  \includegraphics[width = 0.325\linewidth,trim={0 3in 2.5in 3in}, clip]{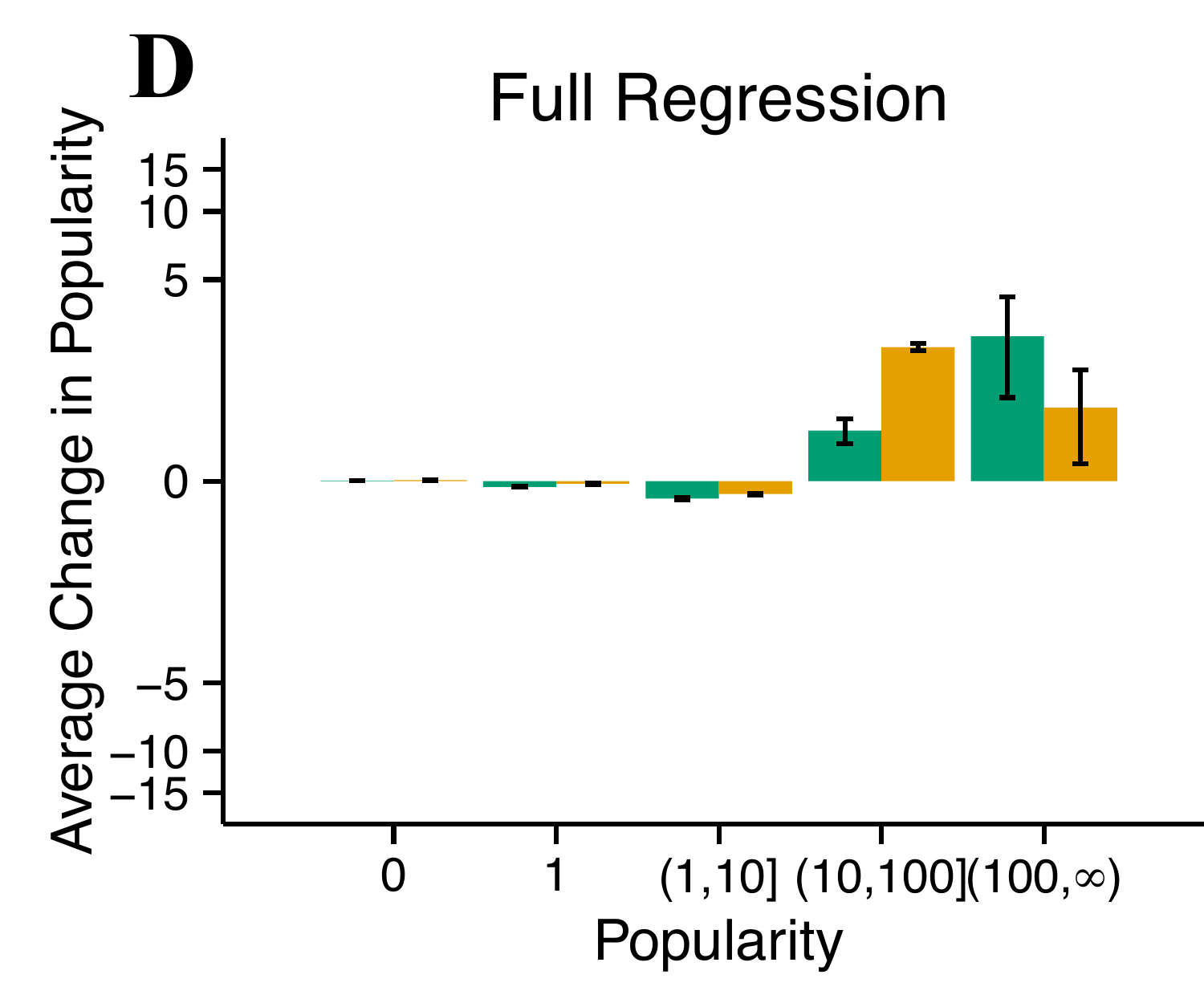}
  \\
  \includegraphics[width = 0.325\linewidth,trim={0 3in 2.5in 3in}, clip]{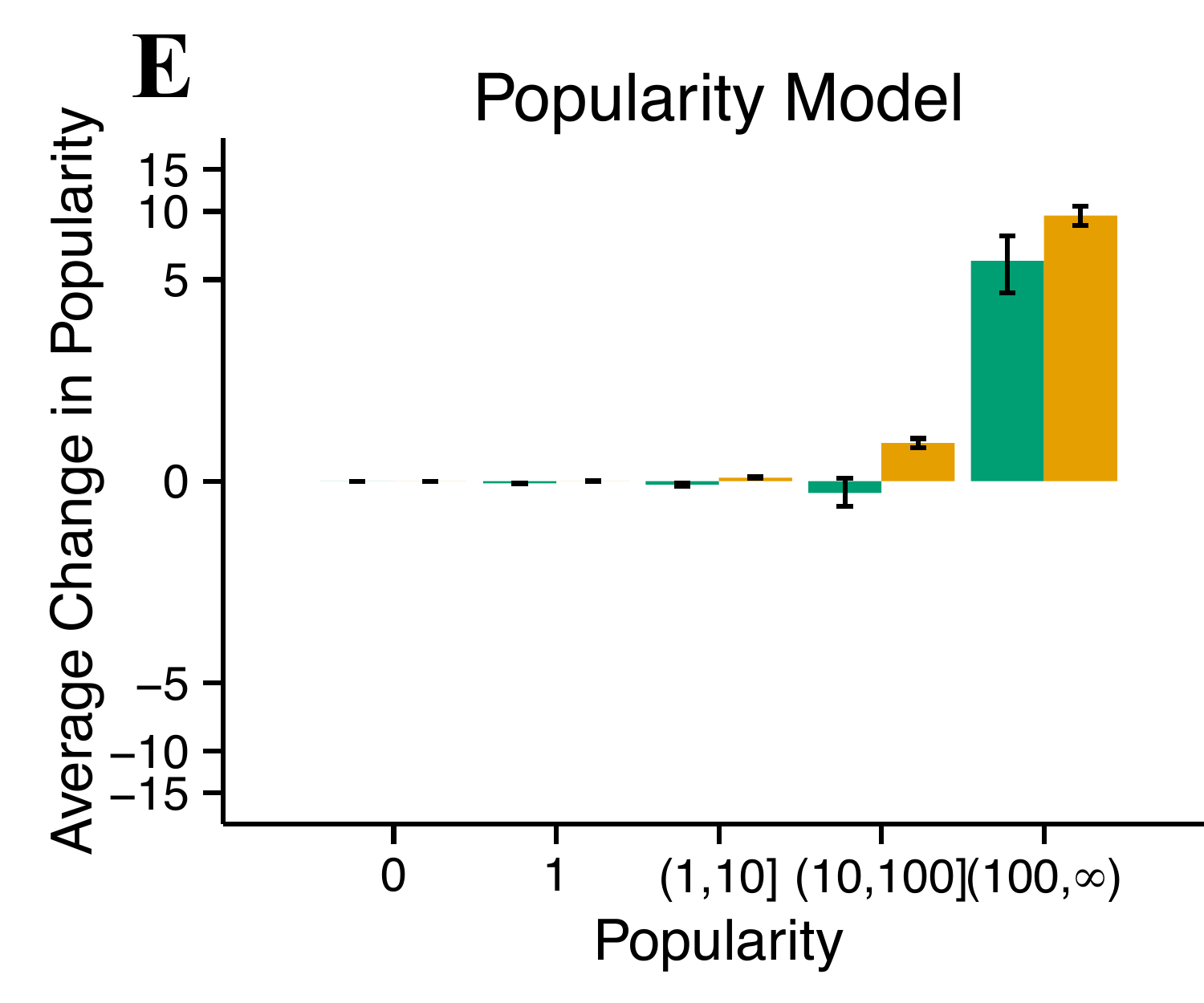}
  \includegraphics[width = 0.325\linewidth,trim={0 3in 2.5in 3in}, clip]{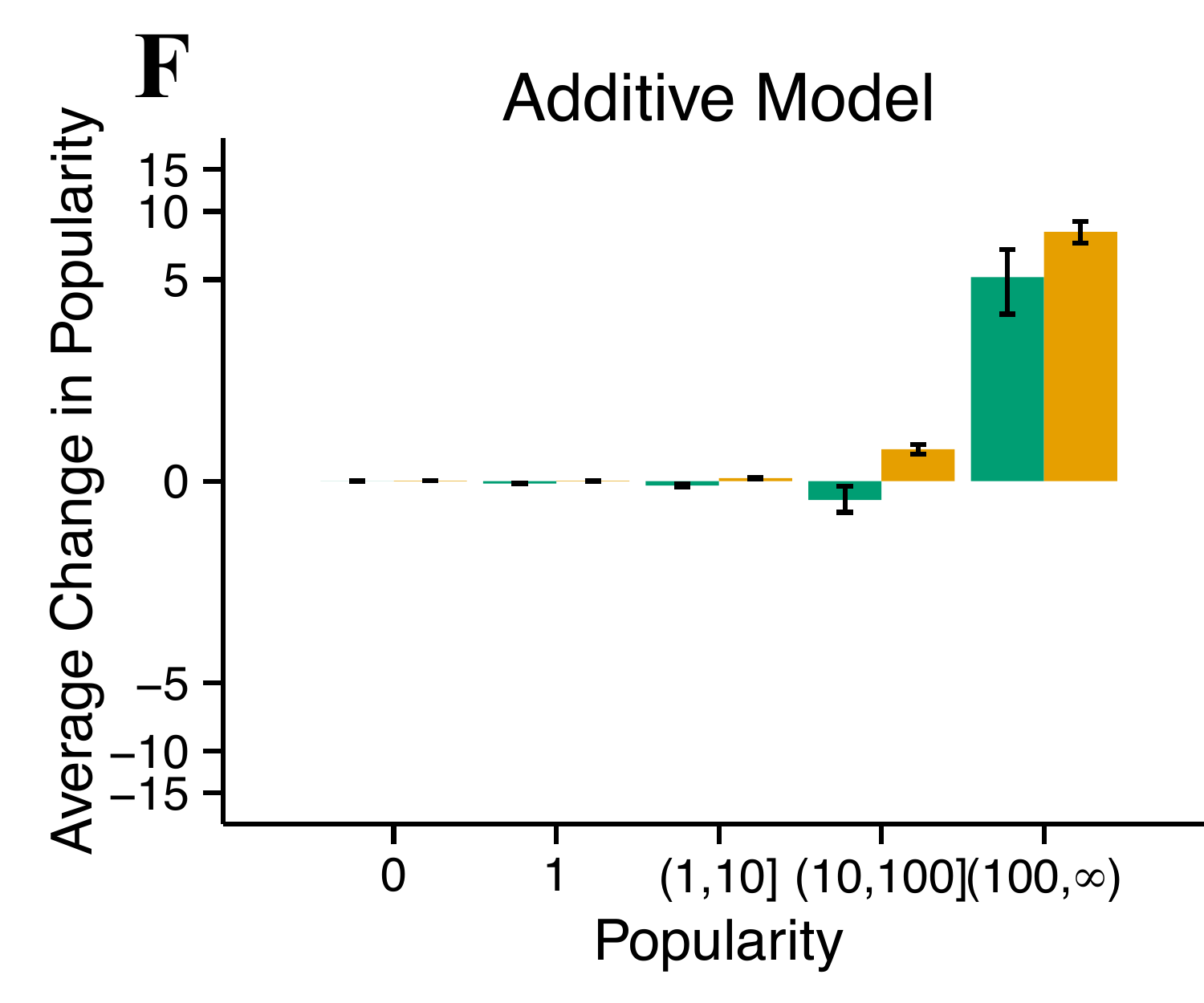}
  \includegraphics[width = 0.325\linewidth,trim={0 3in 2.5in 3in}, clip]{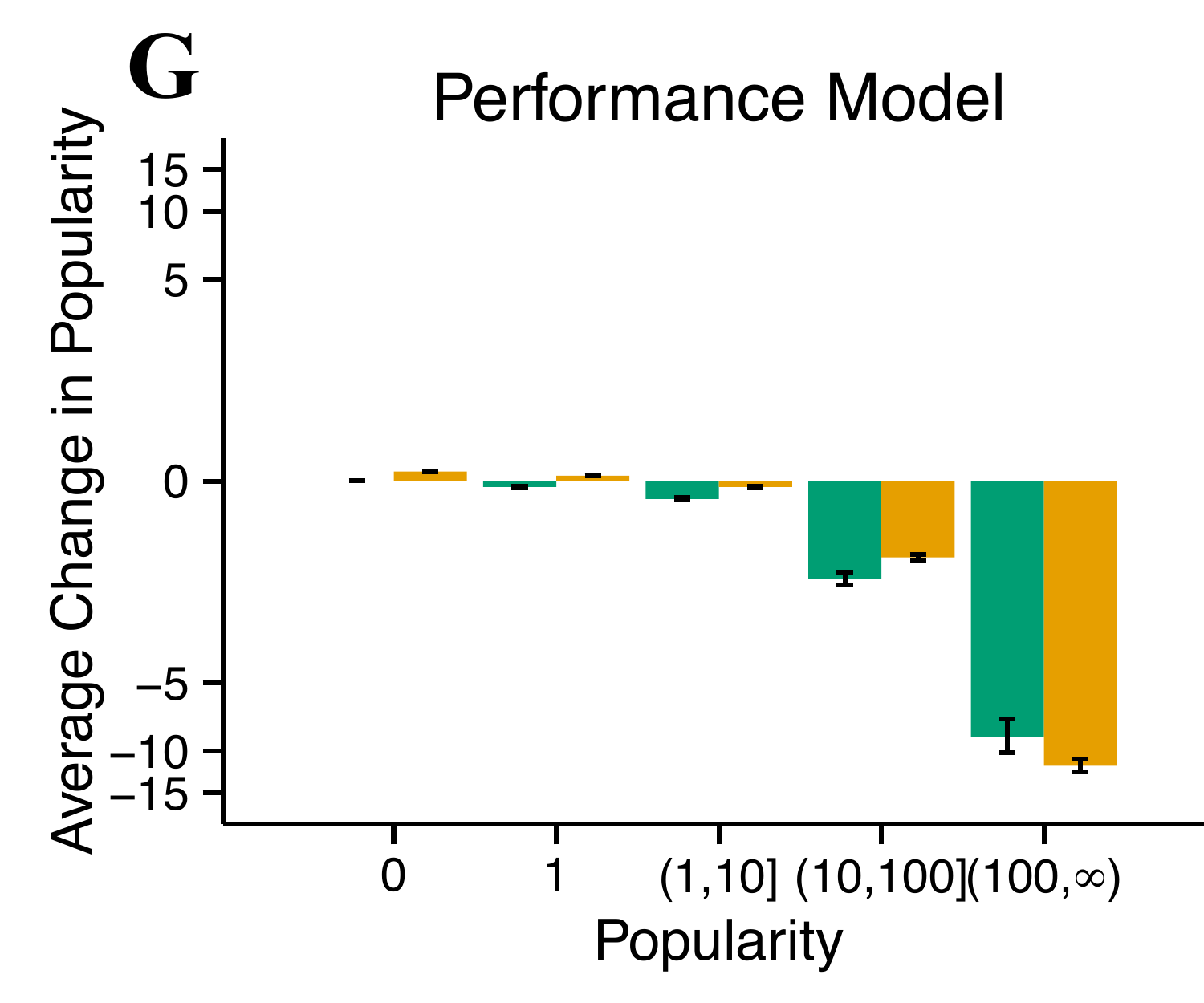}
  \caption{(A) Observational evidence for the multiplicative
    interaction between popularity and performance in determining
    future popularity predicted by the social sampling
    model. (Interaction term $p = 0.012$ in a linear model with fixed
    effects for users and days, and Arellano robust standard errors
    \cite{arellano_computing_1987} to adjust for non-equal variance
    and correlated error terms.)  Error bars are 95\% Gaussian
    confidence intervals of the means.  (B-G) Replications of Panel
    A using predicted gains in mimickers according to each model
    considered, with losses in mimickers taken from the behavioral
    data.  To generate the predictions of each model, we compute the
    expected number of new mimickers each user gets on each day given
    the actual total number of new mimic decisions.  The social
    sampling model provides a better fit to our data than all the
    alternative models.}
  \label{fig:interaction}
\end{figure*}

One striking fact apparent in this dataset is that users are more
likely to mimic popular traders, but only if those traders are
performing well.  While traders on eToro always tend to gain mimickers
when they are performing well and lose mimickers when they are
performing poorly, the magnitudes of these changes are larger when a
trader is more popular.  More specifically, we see that daily changes
in popularity for each user are predicted by a multiplicative
interaction between the past day's performance and popularity (Figure
\ref{fig:interaction}A).  People therefore rely heavily on social information even in the
presence of explicit, public signals about which traders perform best.
%which would be surprising under traditional economic rational models
%of belief formation.  It is also inconsistent with simple heuristic
%models of social influence, in which agents copy others regardless of
%context.  Instead perceived quality strongly modulates the effect of
%social influence.

The social sampling model reproduces the multiplicative interaction
between popularity and performance that we observe in our data (Figure
\ref{fig:interaction}B).  The predictions of the social sampling model
also compare favorably to five alternative models that were chosen to
represent the predictions of simple heuristic and purely rational
models of social decision-making, as well as other plausible
alternatives (see Section ``Model Comparison'' in the appendix for
further model comparisons).  None of these alternatives reproduces the
qualitative form of the popularity-performance interaction (Figures
\ref{fig:interaction}C through \ref{fig:interaction}G).  Models
relying too heavily on social information overestimate the increases
in popularity of already-popular users (Figures \ref{fig:interaction}E
and \ref{fig:interaction}F), and models relying too heavily on
performance underestimate those increases (Figures
\ref{fig:interaction}C and \ref{fig:interaction}G).

\begin{figure*}
  \centering
  \includegraphics[width = 0.3\linewidth, trim={0 1.9in 0in 1.5in}, clip]{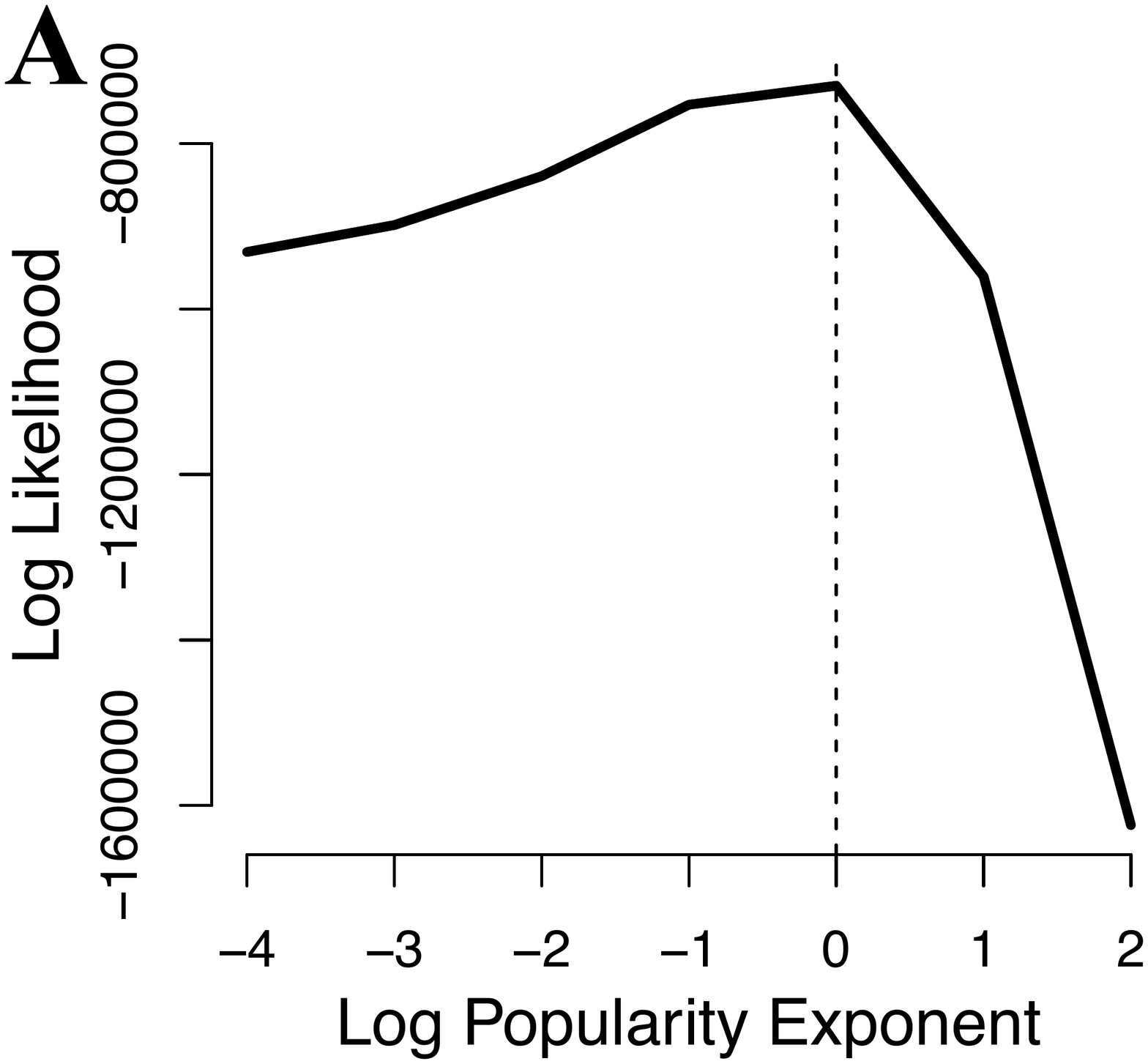}
  \includegraphics[width = 0.6\linewidth, trim={0 4in 0in 3in}, clip]{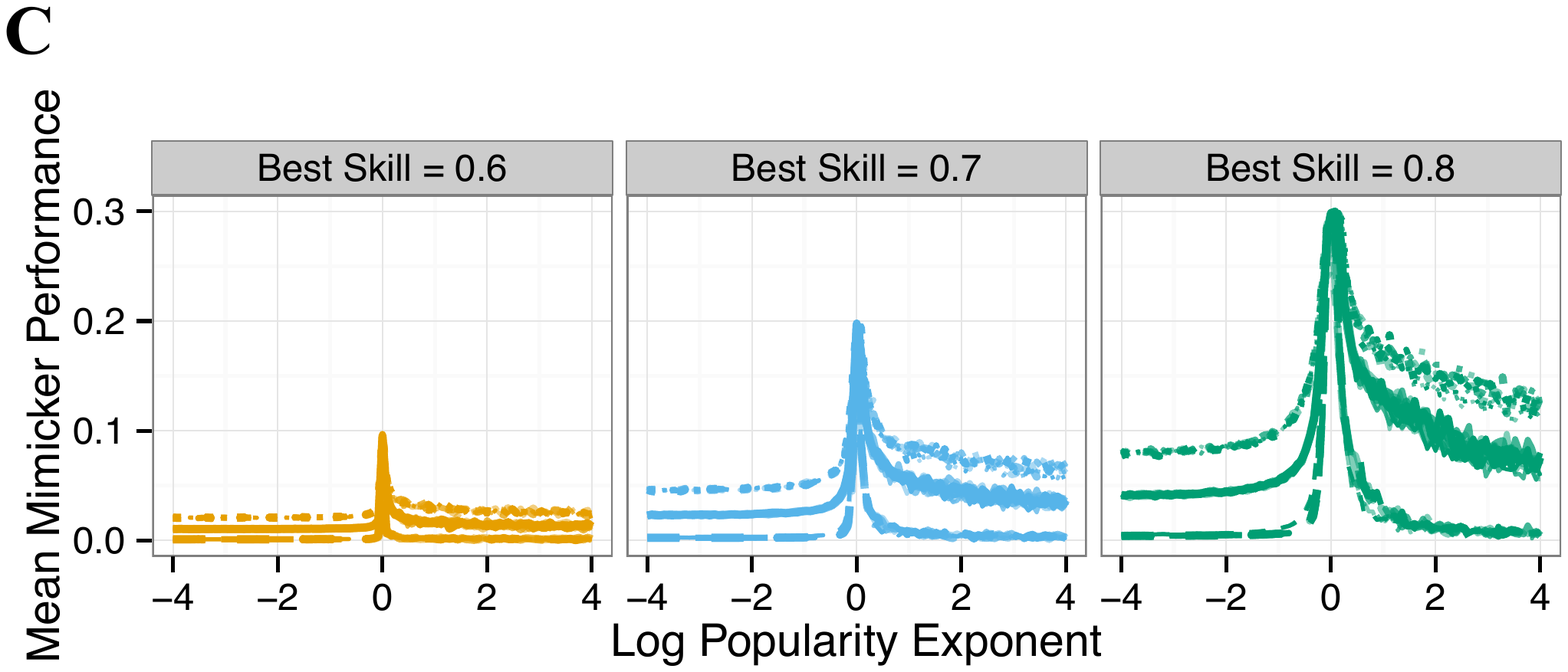}
  \\
  \includegraphics[width = 0.3\linewidth, trim={0 2in 0in 1.5in}, clip]{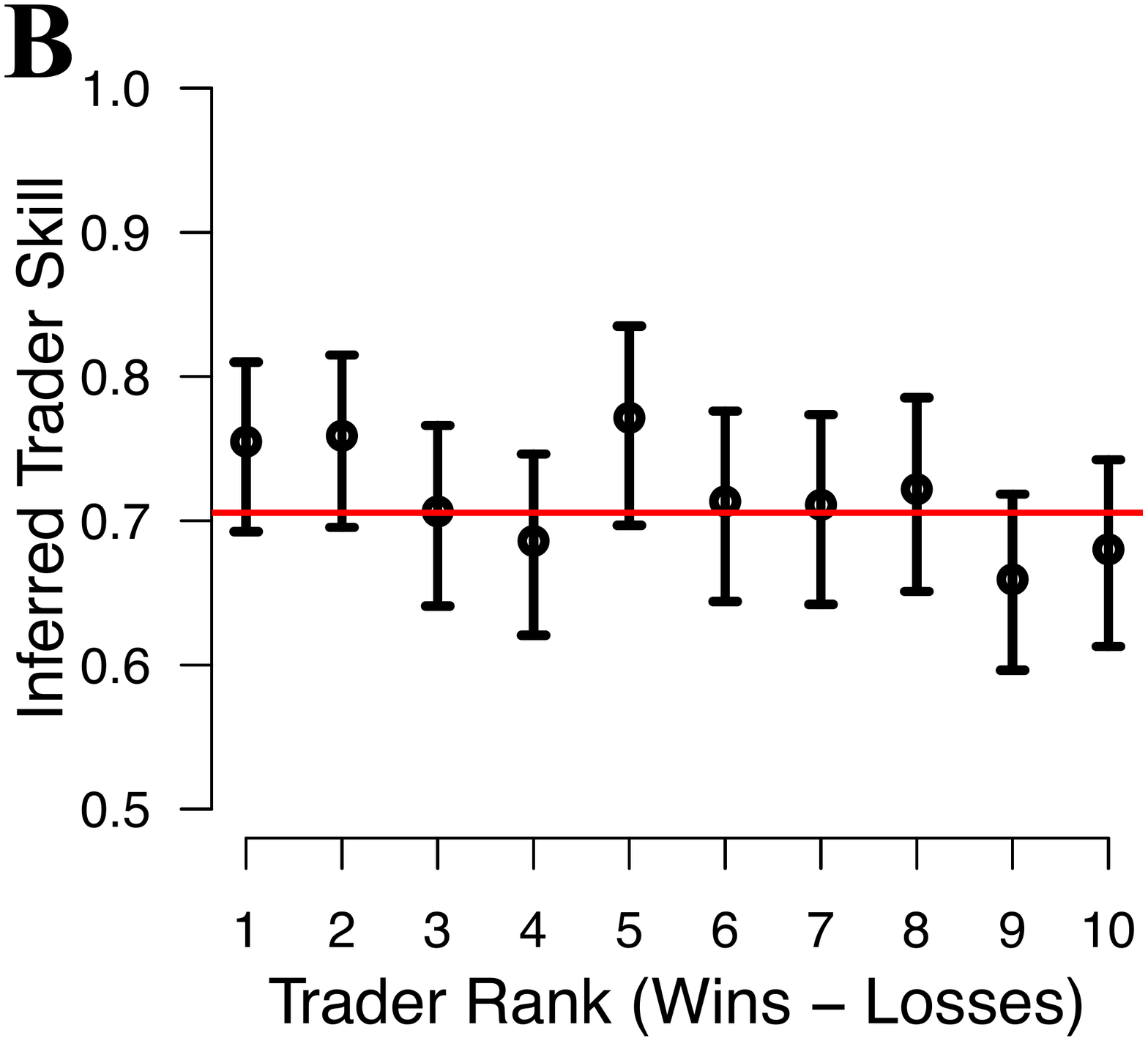}
  \includegraphics[width = 0.6\linewidth, trim={0 4in 0in 3in}, clip]{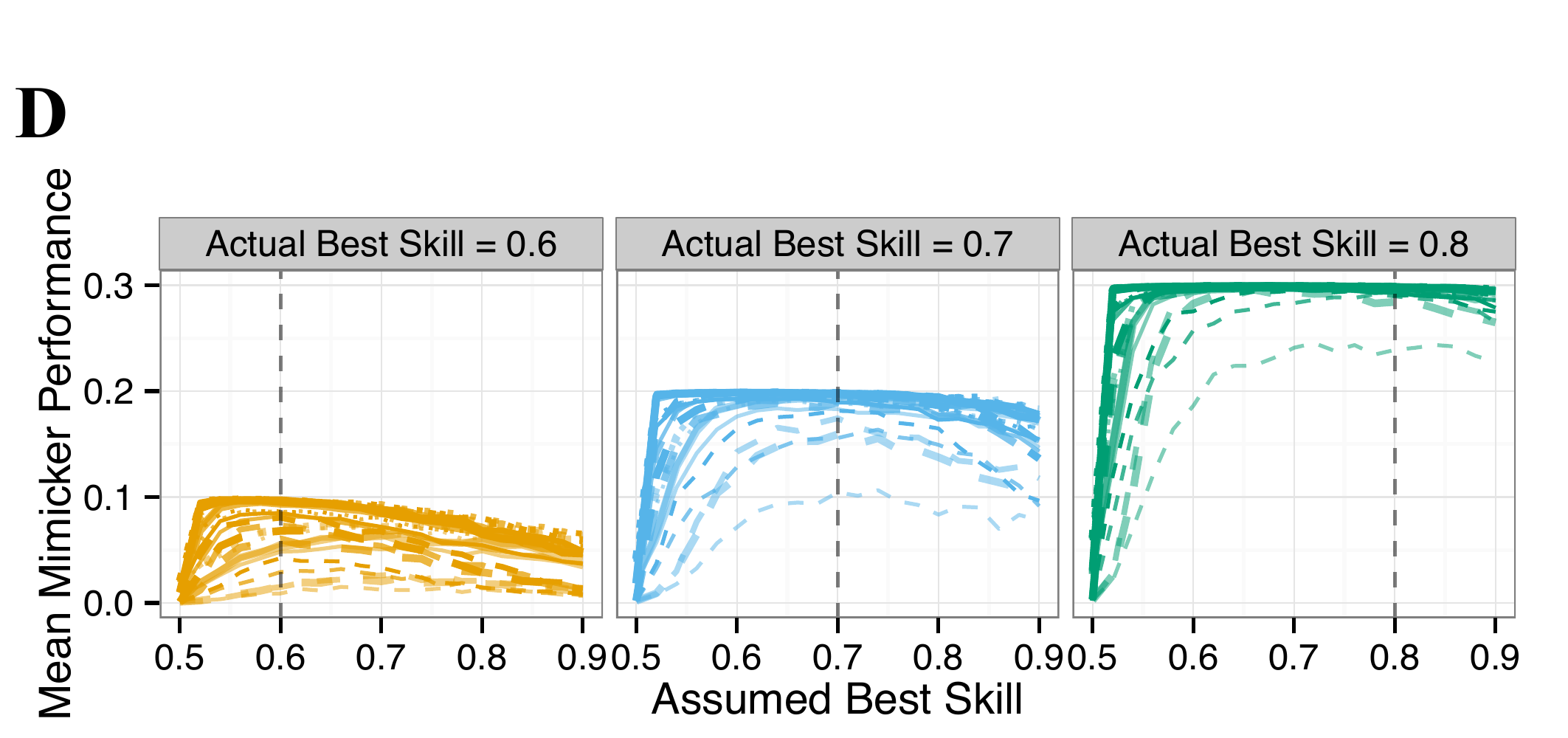}
  \caption{Further evidence from our data for the specific
    mathematical form of the social sampling model, and evidence from
    simulations for how the social sampling model yields collectively
    rational shared belief formation.  (A) Log likelihood values for
    alternative scaling exponents on popularity in the social sampling
    model.  The vertical line indicates the maximum.  A log popularity
    exponent of 0, which corresponds to using popularity linearly as
    in the social sampling model, achieves the best fit.  (B) 95\%
    credible intervals for how often each of the ten best traders in
    our dataset has positive daily return.  The horizontal red line
    indicates the model's inferred $\eta$ parameter.  The fact that
    this line passes through all of the credible intervals indicates
    that the fitted model independently captures this property of the
    data. (C-D) Two idealized simulation experiments.  Each line
    represents a different combination of simulation parameters, with
    the four graphical parameters of the plotted lines (line type,
    width, color, and transparency) each representing a variable.  An
    ``assumed best skill'' value of 0.5 corresponds to decision-makers
    ignoring performance information.  The fact that group performance
    peaks at 0 in Panel C indicates that using popularity linearly, as
    in the social sampling model, is critical to collective
    rationality.  At the same time, the flatness of the curves in
    Panel D indicates that social sampling is otherwise robust to
    deviations in individuals' mental models of the world.}
  \label{fig:model-fit}
\end{figure*}

The Bayesian interpretation of social sampling is also supported by
specific quantitative features of our data.  Rather than using
popularity directly as a prior, as in social sampling, individuals
could hypothetically form a prior based on a superlinear or sublinear
transformation of popularity, $(p_k)^\gamma$.  Figure \ref{fig:model-fit}A shows that
using popularity linearly ($\gamma = 1$) is a better fit to our data.
We also find that the empirically best fitting value of the social
sampling model's $\eta$ parameter is consistent with the parameter's
Bayesian interpretation.  The fitted value of $\eta$ matches what
could independently be expected to be the highest frequency of good
performance signals from any trader in our dataset, even though the
parameter is inferred from users' mimic decisions and not directly
from performance information (Figure \ref{fig:model-fit}B).

But why might a group of individuals use social sampling?  We have
already explained how social sampling in the aggregate yields rational
individual posterior-matched decisions.  Here we note that the
rational mechanism that social sampling collectively implements
closely approximates a well-known single-agent algorithm called
``Thompson sampling''.  Thompson sampling is a Bayesian algorithm for
sequential decision-making that consists of probability matching on
the posterior probability that an option is best at each point in
time.  Thompson sampling has been shown in the single-agent case to
have state-of-the-art empirical performance and strong optimality
guarantees in sequential decision-making
problems \cite{kaufmann2012thompson}.  As long as popularity remains
well-calibrated, social sampling therefore collectively attains these
benefits while avoiding the need for any agent to incur the cost of
computing a full posterior distribution.

Furthermore, we find that social sampling has unique benefits to group
performance compared to nearby models that do not yield a Bayesian
algorithm in the aggregate.  Firstly, using popularity linearly is
critical for the collective rationality of social sampling: it
provides dramatically better group outcomes as compared to even small
deviations from linearity (Figure \ref{fig:model-fit}C).  Hence the
identification of popularity with a Bayesian prior in individuals'
``posterior matched'' choices appears to be necessary for information
from individual decisions to accumulate at the group level in a
collectively rational fashion.  Further simulations also indicate that
incorporating performance information into decision-making is
normatively vastly better than ignoring it, though the specific value
of the $\eta$ parameter is not critical (Figure \ref{fig:model-fit}D).
This result suggests that the benefits of social sampling may be
robust to a group having an inaccurate mental model of the world, at
least as long as individuals in the group are making their decisions
according to some reasonable shared mental model.  Having a shared
mental model allows information to accumulate in a Bayesian fashion,
while having an inaccurate model appears to only lead to mild
inefficiencies.

\section*{Discussion}

These findings support our view of collective intelligence as
distributed Bayesian inference.  The more general utility of this
framework will come from providing a principled constraint on
individual-level mechanisms in modeling collective behavior, and from
providing a rigorous way to relate these individual-level mechanisms
to group-level models.  Looking to the literatures on computer
science, statistics, and signal processing for relevant distributed
inference algorithms is likely to yield a fruitful path towards a new
class of models of human collective behavior.  At the same time,
instances of this new class of models, such as our social sampling
model, will bring novel algorithms to the area of distributed Bayesian
inference.

\section*{Materials and Methods}

\subsection*{Data Processing}
The dataset provided to us by eToro consists of a set of trades from
the eToro website.  The aggregated data we used in our analyses will
be released upon publication.  The entire dataset contains trades that
were closed between June 1, 2011 and November 21, 2013.  Our initial
explorations (which multiple authors engaged in) must be assumed to
have touched all of the data.  When we began systematic analysis, in
order to have a held-out confirmatory validation set for this
analysis, we split the dataset into two years.  We use the first year
of data for all our main analysis.  We then verify that our findings
from these analyses still hold in the second year.  The second year
was almost completely held out after this point, but did witness at
least one major model iteration that was primarily motivated by
theoretical considerations and a lack of fit of a previous model on
the first year of data.  Ultimately, the results on the second year of
data are similar to those on the first year, with the most notable
differences being that the alternative model that relies on social
information alone has more competitive predictive performance on the
second year.
%% , and we did not observe a predicted increase in mean group
%% performance under the social sampling model as compared to the
%% non-social model in the second year.
We also test the robustness of
our main analysis to the specification of the performance metric we
use and the way we parsed our raw data, and we find that our results
are highly robust to these changes.  (See Sections ``Data Processing''
and ``Robustness Checks'' in the appendix for
further details on data processing and robustness analysis.)

\subsection*{Predicting Mimic Decisions}
%% Several aspects of our analysis rely on computing the mimic decisions
%% users on eToro would have been expected to have made according to the
%% social sampling model and alternative models.
Our analysis relied on computing the mimic decisions users on eToro
would have been expected to have made according to the social sampling
model and alternative models.  In all cases we examine aggregations of
these decisions in the form of predicting the total number of new
mimickers each trader on eToro obtains.  More specifically, we predict
the number of new mimickers each user gets on each day given the
performance and popularity of that user (and of every other user) on
the previous day, and given the total number of new mimic events we
observe on those days.

In the social sampling model, decision-makers make decisions
independently, so the probability that a given decision-maker chooses
a specific option $j$ at time $t$ is given by the decision probability
$\theta^{main}_{j,t}$,
$$
\theta^{main}_{j,t} 
= \frac{
  \eta^{r_{j,t}}(1 - \eta)^{1 - r_{j,t}} \cdot (p_{j,t} + \epsilon_t)
}{
  \sum_k \eta^{r_{k,t}}(1 - \eta)^{1 - r_{k,t}} \cdot (p_{k,t} + \epsilon_t)
},
$$ where $\epsilon_t > 0$ is a small smoothing parameter that ensures
all users have some probability of gaining mimickers, and $r_{j,t}$ in
the case of the eToro data simply indicates whether user $j$ has
positive or negative performance on day $t$.  We arbitrarily choose
$\epsilon_t = \frac{1}{M_t}$, where $M_t$ is the number of active
users on day $t$.  The distribution of the number of new mimickers
each user gets on day $t$ will then be given by a multinomial
distribution over the $M_t$ options with parameters equal to the total
number of new mimickers on that day and the vector of probabilities
$[\theta^{main}_{1,t}, \ldots, \theta^{main}_{M_t,t}]$.  Hence the
expected number of new mimickers a particular user $j$ gets on a
particular day $t$ will be $\theta^{main}_{j,t}$ times the total
number of new mimickers on that day.

\subsection*{Alternative Models}
We consider a set of alternative models in order to identify how well
the social sampling model is able to account for structure in the mimic
decisions present in our data compared to alternative plausible
models.  We specify these alternative models in terms of ``decision
probabilities'' analogous to $\theta^{main}_{j,t}$.  These decision
probabilities provide the probability under each model that an
individual will decide to commit to a particular option (or mimic a
particular trader in the case of eToro).  It is possible to specify
all our alternative models in this way because every model we consider
assumes that all decisions are conditionally independent given the
popularity and performance of every option.

The first of these alternatives is a proxy for a probability matching
rational agent model that we call the ``Performance Regression''
model.  This model uses only performance information, and does not
reduce the performance signals to being binary as our social sampling
model does.  The decision probability under this model is
$$
\theta^{perf}_{j,t} \propto  \sigma(\beta_0 + \beta_1 q_{j,t}),
$$ where $\sigma$ is the logistic function and the $\beta_i$ variables
are free parameters.  In combination with our explorations of
different performance metrics (described in Section ``Robustness Checks''
of the appendix), the performance regression allows us to evaluate the
predictive power of using performance information alone to predict
mimic decisions.

The next alternative we consider is an extended regression model that
we call the ``Full Regression'' model.  This alternative consists of a
generalized linear model that includes an interaction term between
popularity and performance.  Such a model could conceivably generate
the multiplicative interaction effects we observe in the eToro data,
but lacks some of the additional structure that the social sampling
model has. The full regression assumes that a decision-maker chooses
to commit to option $j$ with probability
$$
\theta^{full}_{j,t} \propto  \sigma(\beta_0 + \beta_1 q_{j,t} + \beta_2 p_{j,t} + \beta_3 q_{j,t} p_{j,t} ),
$$ where the notation is as above.
The purpose of comparing to this
alternative is to test whether having the additional structure of
including popularity as a prior in the social sampling model lends
additional predictive power, as compared to having a heuristic
combination of popularity and performance.

We also consider a reduction of the social sampling model that does
not use performance information.  We call this model the ``Popularity
Model''.  Under the popularity model the decision probability becomes
$$
\theta^{pop}_{j,t} \propto p_{j,t} + \epsilon_t,
$$ and again we use $\epsilon_t = \frac{1}{M_t}$ as the smoothing
parameter.  Comparing to this preferential attachment model allows us
to understand how much predictive power we get from including
performance information while controlling for the structure of how
social information is used in the social sampling model.  This
preferential attachment model is a canonical simple heuristic model of
social decision-making.

We also consider an alternative model that is a reduction of the
social sampling model that uses only performance information.  This
model, which we refer to as the ``Performance Model'', uses the
decision probability
$$
\theta^{perfm}_{j,t} \propto \eta^{r_{j,t}}(1 - \eta)^{1 - r_{j,t}}.
$$
Since the performance model is the one that would be obtained from
the social sampling model when all options have the same popularity,
this model allows us to predict how decision-makers might behave if
they did not have social information.

Our final alternative model is an additive combination of the
popularity model and the performance model.  This model, which we call
the ``Additive Model'', represents a situation in which some agents
choose whom to mimic based on preferential attachment while other
choose based on performance.  Under this model the decision
probability becomes
$$
\theta^{add}_{j,t} \propto \alpha \frac{(p_{j,t} + \epsilon_t)}{\sum_k (p_{k,t} + \epsilon_t)} + (1 - \alpha)\eta^{r_{j,t}}(1 - \eta)^{1 - r_{j,t}},
$$ where $\alpha \in [0,1]$ is a free parameter and again we use
$\epsilon_t = \frac{1}{M_t}$ as the smoothing parameter.  Comparing to
the additive model allows us to verify that popularity and performance
are combined multiplicatively rather than additively.

\subsection*{Parameter Fitting}
To estimate the parameters of these models we use a maximum likelihood
procedure.  Letting $T$ denote the number of days we observe, letting
$M_t$ denote the total number of users with defined performance scores
on day $t$, and letting $n_{j,t}$ denote the number of new mimickers
user $j$ receives on day $t$, the likelihood of the parameters given
all of the new mimic decisions is
$$
\prod_{t = 1}^T \prod_{j = 1}^{M_t} \left(\theta_{j,t}\right)^{n_{j,t}},
$$ where the $\theta_{j,t}$ is determined by whichever of the models
we are fitting.  To obtain the $\alpha$, $\beta_i$, and $\eta$
parameters in these models we then optimize this likelihood function
using a Nelder-Mead simplex algorithm (in log space for the $\alpha$
and $\eta$ parameters).  We initialize these optimization routines
with values given by grid searches over $[-10, -1, 0, 1, 10]$ for the
social sampling model, the performance regression, the additive model,
and the performance model, and over $[-1,1]$ for the full regression
(the grid search is coarser here since the number of parameters is
larger in this model).

\subsection*{Checking Inferred Parameter Values}
We executed two tests to provide further evidence for the specific
parametric form of the social sampling model.  We first examined
whether using a scaling exponent on popularity would have led to a
better fit to our data.  This modification gives the following form
for the decision probability:
$$ \theta^{scaled}_{j,t} \propto \eta^{r_{j,t}}(1 - \eta)^{1 -
  r_{j,t}} \cdot (p_{j,t}^\gamma + \epsilon_t),
$$ where the scaling exponent $\gamma$ is now another free parameter.
To examine the fit of the model under various values of this scaling
exponent, we simply fix $\gamma$ equal to each value in a coarse grid,
then find the best maximum likelihood value under that $\gamma$ value
using the likelihood function given in Section ``Parameter Fitting''
above.  These maximum likelihood values are plotted in Figure \ref{fig:model-fit}A.

We also examined the inferred value of the $\eta$ parameter in the
social sampling model ($\gamma = 1$).  $\eta$ represents the expected
value of committing to the best option, or in the case of the eToro
data, the skill of the best trader as measured by the expected
proportion of performance signals that will be greater than zero.  To
arrive at plausible actual values for what the skill of the ``best
trader'' on eToro might be, we first rank all traders according to the
amount of evidence for their success.  For this ranking we use an
aggregated single-day net profit values.  We achieve the ranking of
the traders by taking the total number of days each trader had
positive single-day profit and subtracting the total number of days
those traders had negative single-day profit.  This metric
simultaneously considers both the total amount of positive or negative
evidence for trader skill in addition to the proportion of positive
evidence. For each of the top ten traders according to this metric, we
then compute a 95\% Bayesian confidence interval (under a uniform
prior) of the probability that those users will achieve positive
performance, assuming a Bernoulli model.  These confidence intervals
are then plotted in Figure \ref{fig:model-fit}B along with the actual inferred ``best
skill'' $\eta$ parameter.

%% \section{Measuring Predicted Group Performance in Real Data}
%% In order to assess the broader implications of the social sampling
%% model, we compared the predicted mimic decisions from the social
%% sampling to those of the performance model in terms of group
%% performance.  To generate these predictions, we use the eToro data to
%% compute the number of new mimickers each trader is expected to get on
%% each day under the social and non-social models, and we then evaluate
%% the single-day return from those new mimic decisions.  To compute the
%% single-day returns and the ``ROI of the group portfolio'', we computed
%% the daily ROI of each trader on eToro resulting from both closed
%% trades and liquidated open trades.  The daily ROI for a trader on a
%% particular day is defined as the total profit that trader made on that
%% day divided by the total amount invested.  We then take a weighted
%% average of the ROI values of all users on each day, with the weights
%% given by the number of new mimickers those users get under each model.
%% The differences between the distributions in Figure \ref{fig:simulations} are jointly
%% statistically significant.  The nonparametric Mann-Whitney $U$ test
%% produces a two-sided $p$-value of 0.0011.  The increase in mean
%% performance occurs via a reduction in loss since both the mean social
%% trader and the mean non-social trader on eToro have slightly negative
%% expected performance.

\subsection*{Idealized Simulations}
To perform our idealized simulation, we implement the environment
assumed in a theoretical justification of our model (described in
Section ``Social Sampling Model Specification'' of the appendix).  In
these simulations, there are $M$ options that $N$ agents can choose to
commit to on each of $T$ steps.  Each of these options generates a
reward, either 0 or 1 on each round, with the rewards chosen according
to independent Bernoulli draws.  We suppose that committing to a
decision has a cost of 0.5.  Then $M - 1$ of these options have
expected return 0, while one option---the ``best option''---has
positive expected return.  In our simulations the best option has
Bernoulli parameter $\eta^*$, which can be different from the agents'
assumed $\eta$.  At time $t$, the agents are able to observe the
decisions made in round $t - 1$, as well as the reward signals from
round $t - 1$.  The agents in these simulations make their decisions
according to the social sampling model strategy, in some cases with
alternative scalings on popularity.  In each round, every agent first
selects an option to consider with probability proportional to
$p_{j,t}^\gamma + \frac{1}{M}$, where the notation is as above.  Each
agent then chooses to commit to the option that agent is considering
with probability $\eta^{r_{j,t}}(1 - \eta)^{1 - r_{j,t}}$.  When
$\gamma = 0$ this process becomes the performance model.  When $\eta =
0.5$ this process becomes the popularity model.

%We conduct three sets of simulations.
We conduct two sets of simulations.  The first set examined the impact
of alternative $\gamma$ scaling exponents.  For this experiment we
look at the average reward in the final round achieved by agents who
committed to some option in that round, which we call the ``Mean
Mimicker Performance''.  The results of these simulations are shown in
Figure \ref{fig:model-fit}C.  Each line in this figure represents a
different combination of simulation parameters.  We look at all
combinations of $N \in [1000, 5000, 10000]$, $M \in [5, 10, 100]$, $T
\in [100, 500, 1000]$, $\eta^* \in [0.6, 0.7, 0.8]$. For this
simulation experiment, we assume $\eta = \eta^*$.  Each data point is
an average over 500 repetitions for that particular combination of
simulation parameters.  The panels of the figure are separated by
$\eta^*$ value.  Line color also indicates $\eta^*$, line size
indicates $N$, line type indicates $M$, and transparency indicates
$T$.

The second set examined the impact of agents having an inaccurate
model of the world.  We again look at mean mimicker performance in the
final round of each simulation.  Here, though, we fix $\gamma = 1$ and
consider $\eta$ values ranging from 0.5 to 0.9, independent of the
value of $\eta^*$.  (Note again, $\eta = 0.5$ corresponds to ignoring
performance information.)  The results of these simulations are shown
in Figure \ref{fig:model-fit}D.  Each line in this figure represents a different
combination of simulation parameters, and the parameter sweep is over
the same space as the first set of simulations.  The panels and line
characteristics are determined as in the first set of simulations.

\subsection*{Methodological Limitations}
One aspect that our methodology cannot identify is the role that the
eToro interface plays in shaping user behavior on the site.  We
observe evidence for a multiplicative interaction between popularity
and perceived quality even at low levels of popularity (see Section
``Possible Confounding Factors'' of the appendix for details), which
suggests that the social sampling model holds independently of the
encouragement of the interface.  However, the interface is likely
contributing to the effect at high levels of popularity.  One
plausible way social sampling could be implemented on eToro is by
users sorting others by popularity, then choosing to commit primarily
based on perceived objective quality.  Regardless, the mere fact that
users find this interface intuitive and useful supports social
sampling as a natural decision-making strategy (see Section
``Anecdotal Evidence'' of the appendix for anecdotal reports), and the
site designers may explicitly or implicitly have tuned the interface
to natural human behavior.
%% Whether the ultimate
%% outcome is an interface that conforms to natural behavior, or an
%% interface that exaggerates a natural behavior, is an interesting open
%% question.

\subsubsection*{Acknowledgments}
This research was partially sponsored by the Army Research Laboratory
under Cooperative Agreement Number W911NF-09-2-0053 and is based upon
work supported by the National Science Foundation Graduate Research
Fellowship under Grant No. 1122374. Views and conclusions in this
document are those of the authors and should not be interpreted as
representing the policies, either expressed or implied, of the
sponsors.

\bibliographystyle{unsrt}
\bibliography{etoro}

\appendix

\section*{Appendix}

\renewcommand{\thefigure}{S\arabic{figure}}
\renewcommand{\thetable}{S\arabic{table}}

\setcounter{figure}{0}
\setcounter{table}{0}

\subsection*{Data Source}

We received our data from a company called eToro.  The data was
generated from the normal activity of users of their website,
etoro.com.  The two main features of the eToro website during the time
our dataset was being collected were a platform that allowed users to
conduct individual trades and a platform for finding and mimicking
other users of the site.  We will refer to the site's users
interchangeably as either users or traders---having these two terms
will ultimately reduce the ambiguity in some of our descriptions.  The
internal algorithms and the website design have changed over time, but
the following description represents to the best of our knowledge the
main contents and features of the website during the time period of
our data.

The eToro website includes basic functionality for use as a simple
trading platform.  This platform allows users to enter long or short
positions in a variety of assets.  Entering a long position simply
consists of buying a particular asset with a chosen currency.
Entering a short position consists of borrowing the same asset to sell
on the spot, with a promise to buy that asset at a later time.  Taking
a long position is profitable if the price of the asset increases,
while taking a short position is profitable if the price of the asset
decreases.  Users can also enter leveraged positions.  A leveraged
position is one in which an user borrows funds in order to multiply
returns.  Leveraged positions have more risk because users will lose
their own investment at a faster rate if the price of the asset
decreases.

At the time our data was collected, eToro focused on the foreign
exchange market, so the trading activity mainly consisted of users
trading in currency pairs---buying and selling one currency with
another currency.  However, users were also able to buy or sell other
commodities such as gold, silver, and oil, and eventually certain
stocks and bundled assets.  The average amount of money invested in
individual trades on eToro was about \$30, and, after accounting for
leverage, individual trades on average result in about \$4000 of
purchasing power.  These amounts are small compared to the trillions
of dollars traded daily in the foreign exchange and commodity
markets\footnote{According to the Bank for International Settlements'
  2013 ``Triennial Central Bank Survey'', the foreign exchange market
  (in which most of the trading on eToro occurs) has a daily trading
  volume of trillions of USD.}, so individual traders are unlikely to
have substantial market impact with their trades.

Besides providing a platform for individual trading, eToro also offers
users the ability to view and mimic the trades of other users on their
website.  To be clear in our terminology, when referring to one user
mimicking another user, we will call the first user the ``mimicking
user'' and the second the ``target user''.  When referring to a
specific mimicked trade, we will refer to the original trade as the
``parent trade'' and the copy as the ``mimicked trade''.  eToro refers
to ``mimicking'' as ``copying'', and ``mimickers'' as ``copiers''.  We
use the term ``mimic'' rather than ``copy'' so that we can reserve the
word ``copy'' for social influence due to information about
popularity, as in ``copying the crowd in making decisions about whom
to mimic''.  eToro also offers an option to ``follow'' users without
``mimicking'' them.

While there is functionality for copying individual trades on eToro,
we focus on the website's functionality for mirroring all the trades
of specific users.  Mirroring works as follows.  First, a mimicking user
allocates funds that will be used for mirroring a target user.  The
mimicking user's account then automatically executes all of the trades
that the target user executes.  The sizes of these trades are scaled
up or down according to how much money the mimicking user has allocated
as funds for that mimic relationship.  When beginning a mimic
relationship, the mimicking user can specify either to only mimic new
trades of the target user or to also open positions that mirror all
the target user's existing open positions.  When a user stops mimicking
a target user, the mimicking user can choose to either close all the
open copied trades associated with that relationship or to keep those
trades open.

There are certain limitations that eToro places on mimic trading.  For
example, users can mimic no more than 20 target users with no more than
40\% of available account funds allocated to a single target user.
Users can also make certain adjustments to their copied trades.  For
example, mimicking users can close a trade early or adjust a trade's
``stop loss'' amount.

eToro also offers an interface to assist users in finding traders to
mimic.  The central feature of this interface at the time our data was
collected was a tool that presented a list of other users on the site.
This list could be sorted either by the number of mimickers those users
had or by various performance metrics, such as percentage of
profitable weeks or a metric called ``gain''.  In a separate part of
the site, users also had realtime or near realtime access to details
of individual trades being executed by other users of the site.

In addition to searching for basic information using these tools, the
website also allows users to view more detailed profiles of other
traders on the site.  These profiles present information such as the
number of mimickers the user has had over time, the ``gain'' of the user
over time, and information about opened and closed trades.

\subsection*{Data Processing}

Each entry in the dataset we received includes a unique trade ID, a
user ID, the open date of the trade, the close date of the trade, the
names of the particular assets being traded, the amount of funds being
invested, the number of units being purchased, the multiplying amount
of leverage being used to obtain those units, the open rate of the
pair of assets being traded, the close rate of that pair, and the net
profit from the trade.  For entries associated with copied trades,
there is additional information.  For individually copied trades, the
parent trade ID is included.  For trades resulting from mimic
relationships between users, ``mirror IDs'' are included in addition
to parent trade IDs.  A mirror ID is an integer that uniquely
identifies a specific mimic relationship.  When a user begins to mimic
another user, a new mirror ID for that pair is created.

In order to study the relationship between previous popularity,
perceived quality, and the mimic decisions of users on eToro using
this dataset, we first had to extract the popularity and the
performance of each user on each day.  For our main analysis, to best
match the statistics that the eToro interface presented to users, we
defined performance as investing performance measured as average
return on investment (ROI) from closed trades over a 30-day period.
In this computation, the ROI for a trade is determined by the profit
generated from the trade divided by the amount withdrawn from the
user's account to make the trade.  If on a particular day a user did
not make any trades in the previous 30 days, the performance of that
user is not defined and the user is removed from the analysis for that
day.  This exclusion criterion removes inactive users from the
analysis.

We use ROI rather than a risk-adjusted performance metric because the
most prominent performance metric presented to users on eToro is what
eToro calls ``gain''.  eToro states this metric is computed using a
type of ``modified Dietz formula'', an equation closely related to
ROI.  Thus ROI should better capture the perceived objective quality
of mimicking each user, though perhaps not true underlying quality.
We later test the robustness of our conclusions to our choices of
performance metric, rolling window, and use of only closed
trades. (See Section ``Robustness Checks'' below.)

We estimated the number of mimickers each user had on each day from the
``mirror IDs'' present in the data we received.  We first identify
which two users are participating in each mirror ID, and we then
determine the duration of the mimic relationship between those two
individuals as beginning on the first date we observe that mirror ID
and ending on the last date we observe that mirror ID.  From these
time intervals we can then estimate the number of mimickers each user
has on each day, as well as when each mimic relationship begins.

We conduct our main analysis using approximately one year of data,
from June 1, 2011 to June 30, 2012.  However, we skip the first month
of data when analyzing changes in popularity so that we are only using
the period of time for which we have accurate estimates of previous
popularity and changes in popularity.  We also do not analyze changes
in popularity that occur over weekends.  Only a small percentage of
trades occur on weekends since trading on eToro is closed on Saturdays
and opens late on Sundays.  Since the way we measure changes in
popularity depends on having observed trades, days on which there is
little to no trading can lead to inaccurate estimates of changes in
popularity.

The data frames that we ultimately use for our statistical analyses,
modeling, and model predictions are then constructed as follows.  Each
user is given a row in the data frame for each day on which that user
had any trades in the previous 30 days.  The columns associated with
this row are the performance score of the user (the expected ROI from
closed trades from the previous 30 days), the number of mimickers the
user had on the previous day, the number of new mimickers that user
gained on that day, and the number of mimickers that the user lost.

%For the purposes of evaluating the broader implications of the social
%sampling model,
For the purpose of evaluating the robustness of our conclusions to the
performance metric we use, we also include two additional columns for
each row, i.e. for each user-day pair. The first additional column is
the actual amount of funds invested in new trades on that day by that
user.  This ``actual amount'' is the amount in USD invested in each
trade initiated on that day after accounting for the leverage the
trader used in those trades.
%% We use this leveraged amount for evaluating our
%% model's broader implications rather than the amount directly withdrawn
%% from the user's account in order to arrive at a more principled
%% performance metric. In our analyses of user decisions we use a
%% performance metric that should approximate the signals of quality
%% shown to users of the site, and hence we use the amount directly
%% withdrawn from the users' accounts rather than the leveraged amount.
%% However, for the purpose of better evaluating the actual performance
%% of the different mimicking strategies, we attempt to use the best
%% performance metric we can.

The second additional column we use contains the sum of the realized
and unrealized gains and losses from all new trades each user made on
each day.  To obtain the realized profit or loss for a user on a
particular day, we simply sum the profit from each trade that the user
opened on that day and subsequently closed on the same day.  To obtain
the unrealized profit or loss, we take each trade that the user opened
on that day but did not close.  We then compute the unrealized profit
or loss of those trades as the profit or loss that would have resulted
from those trades being closed at the end of the day.  To obtain the
close rates for these trades, we use an external database of foreign
exchange rates for all the currency pairs, and for other assets (whose
close rates we don't have from an external database) we use the last
observed rate on that day from the eToro data.

\subsection*{Data Limitations}

Large observational datasets frequently contain flaws such as
inconsistencies or missing data.  With regard to missing data, we know
that we were not able to receive every trade closed on eToro during
our observation period.  Our dataset was collected on a rolling
basis, and there were certain periods of time that a lost connection
resulted in entire days missing.  There appear to be 30 such days of
data missing, which is about 4\% of all days during our observation
period, and likely contains about 2\% of the total number of trades
closed during that time.  However, this missingness should not
substantially affect our results because the way we estimate
popularity is relatively robust to missing data, and because the
relative performance rankings of users should be similar after
removing entire days.

There is also evidence that we may be missing additional trades beyond
the trades on these missing days.  There are certain trades that we
know must be copies for which we have not observed the original parent
trades.  The reason for this could be that certain trades are lost at
random, or that these missing individual trades are actually from the
entire days we are missing and the copies were closed early
(e.g. because of a stop loss), or any number of other reasons.  The
percentage of trades that we have direct evidence for being missing in
this way is about 1\% of all trades.  Moreover, the mimicked trades we
do have indicate that almost all of these missing trades are either
near zero profit or unprofitable.  Fortunately, since we have observed
the copies of these missing parent trades, we are able to impute these
missing trades.

This evidence we have that there may be certain individual trades
missing introduces the possibility there are additional trades from
users with no mimickers that are missing without a trace.  However, we
can bound the amount of data that could be missing since trade IDs
appear to be assigned sequentially.  Ignoring the 1,000 smallest trade
IDs we observe, which were from long-running open trades from before
the beginning of our observation period, we observe approximately 92\%
of the remaining possible trade IDs.  Given the 3\% of data we know is
missing, this larger number means we could be missing an additional
5\% of trades without any trace.  However, this amount is necessarily
a loose upper bound, and the additional amount missing could be far
less if trade IDs are skipped for other reasons.  For example, there
must be trades that were still open at the end of our observation
period and hence were not included as closed trades in our data.
While there is little we can do about the possibility of these missing
trades, it is important to note that if these missing trade follow the
pattern of the missing trades we can reconstruct (i.e. missing trades
tend to be unprofitable trades), then their missingness cannot be
causing the effects we observe.  Assuming that missing trades from
users with mimickers typically have at least one copy remaining in our
data, then we should be able to get accurate performance estimates for
traders with higher popularity, and we should overestimate the
performance of low popularity traders.  This overestimation of the
performance of low popularity traders can only be weakening the
effects we identify.  Furthermore, since our popularity estimates are
only based on the dates of the first and last trades within each mimic
relationship, our popularity estimates should not be greatly affected
by these missing data points.

Another limitation of our dataset is occasionally inconsistent column
data.  We find 1,170 trades (about 0.001\%) have negative amounts
invested, 182 trades (about 0.0002\%) have closed dates occurring
before open dates, and 6,470 trades (about 0.007\%) have profit fields
that are inconsistent with the given units invested and rates
observed.  We attribute these inconsistencies to bugs or database
errors from eToro's code base.  We are also not able to reconstruct
the exact relationship between the amounts initially invested in each
trade and the units purchased, though the relationship we presume
between these columns achieves 10\% relative error on about 83\% of
the trades and 50\% relative error on about 96\% of the trades.  It is
likely that the relationship between these columns relies on data we
have not observed, such as ``stop loss'' and ``take profit'' amounts
that users can specify to limit their risk.  To test for robustness to
these inconsistencies, we compute the amount invested by each user and
the profit made in multiple ways, with each way relying on different
data columns, and we check if our conclusions hold using each of these
data parses.

\subsection*{Imputing Missing Trades}

As discussed above, in our dataset there are a small number of
observed mimicked trades that lack parent trades in the dataset.
Since these missing parent trades are predominantly unprofitable, and
since overestimating the performance of popular traders could
substantially bias our results, we developed a method for recovering
these missing parent trades.  Certain fields of mimicked trades
(including the open dates, close dates, assets traded, trade
directions, and associated open and close rates) are typically almost
identical to the fields of their associated parent trades.  These
similarities allow us to recover the direction of profit of these
missing trades with high reliability.  However, the initial amounts
invested and the units purchased in missing parent trades are more
difficult to infer.  To estimate the amounts invested in each missing
parent trade, we use the fact that the ratios between the units
invested in mimicked trades and the units invested in their associated
parent trades are relatively stable for a particular mirror ID.
Specifically, we use the following procedure.  We first, for each
mirror ID, compute the median ratio between the units invested in each
of the observed mimicked trades associated with that mirror ID and the
units in each of those trades' parents.  For a particular missing
parent, we then find all of the mimicked trades in our data with this
missing trade as a parent.  We then gather what the units invested in
the parent trade would be according to each of the median ratios
associated with those copies, and we take the median of those unit
values.  We use this final quantity as the number of units we presume
were purchased in the original trade.  We finally compute the amount
of funds invested in each missing trade and the ultimate profit made
from each of those trades from the inferred units purchased and the
open and close rates of a single observed mimicked trade.  We conduct
our main analysis using these imputed trades, and we test for
robustness by verifying that our conclusions hold on the raw data as
well.

\subsection*{Data Analysis}

In this section we provide more details of the statistical evidence
that there is a positive multiplicative interaction between previous
popularity and performance in determining future popularity.  To
support this hypothesis we perform both a simple analysis using
ordinary linear models and a more robust analysis that accounts for
dependence between data points and individual user-level effects.  Our
model comparisons described in the subsequent sections provide further
evidence for this interaction effect.  For the following analyses, we
use the data format as described at the end of the ``Data Processing''
section above, and we use two-sided hypothesis tests for all
$p$-values.

\begin{table*}[h]
  \centering
  \caption{Results from an ordinary linear regression.}
  \begin{tabular}{|l|l|l|}
    \hline
    \textbf{Independent Variable} & \textbf{Coefficient} & \textbf{$p$-value} \\ \hline
    Intercept   & 8.226e-03   & \textless 2e-16 \\ \hline
    Popularity  & 6.499e-03   & \textless 2e-16 \\ \hline
    Performance & 1.720e-02   & 3.01e-06        \\ \hline
    Interaction & 4.748e-02   & \textless 2e-16 \\ \hline
  \end{tabular}
\end{table*}

Initially, we perform a basic statistical analysis assuming that the
observed changes in popularity within each user and within each day
are conditionally independent of each other given previous popularity
and performance.  That is, this first analysis assumes that the only
pieces of information that affects mimic decisions are popularity and
performance information, and there are no systematic preferences for
particular users and no systematic differences in changes in
popularity on particular days.  The results of this ordinary linear
regression using all of the active users on each day are given in
Table S1.  The dependent variable in this regression is the change in
popularity of each user on each day computed as the user's popularity
on that day minus the user's popularity on the previous day.  The
independent variables are the performance scores of each user computed
from the previous 30 days, the number of mimickers that the user had on
the previous day, and an interaction term multiplying these two
values.  The results of this regression indicate that there is a
significant positive interaction effect between previous popularity
and previous performance in relation to changes in popularity.

\begin{table*}[h]
  \centering
  \caption{Results from fixed effects model with robust standard errors.}
  \begin{tabular}{|l|l|l|}
    \hline
    \textbf{Independent Variable} & \textbf{Coefficient} & \textbf{$p$-value} \\ \hline
    Popularity   & 1.3301e-03   &  0.4007003 \\ \hline
    Performance  & 1.0583e-02  &  0.1814710 \\ \hline
    Interaction & 4.3830e-02   & 0.0115439 \\ \hline
  \end{tabular}
\end{table*}

For a more robust statistical analysis, we used a fixed effects model
with Arellano robust standard errors \cite{arellano_computing_1987}.
We included fixed effects for users and days.  Including fixed effects
for users and days allows us to rule out the possibility that the
effects we observe are driven by a few peculiar users or days, and the
robust standard errors correct the $p$-value for non-equal variance and
correlated error terms.  Besides these changes, the models are the
same as the previous ordinary linear models.  The results of these
fixed effect models are given in Table S2.

\subsection*{Social Sampling Model Specification}

We now provide a formal statement of the ``social sampling'' model
that we propose to account for the statistical effects we observe, and
the details of our Bayesian interpretation of this model.  We first
describe a general, idealized version of the model, and we then
describe how we apply this model to our data.

We assume discrete time, a set of $N$ decision-making agents, and a
set of $M$ possible options that these agents can choose between.  We
assume that the decision-making agents exist in an environment in
which each option $j$ is associated with a particular distribution
over rewards, $P(r_{j,t})$. At each point in time $t$, each option
generates a single binary reward signal $r_{j,t} \in \{-1,1\}$
according to $P(r_{j,t})$.  We suppose that the decision-making agents
assume that one option is the ``best'' option $j^*$.  We further
suppose that the agents assume the distribution of the best option is
$P(r_{j,t} = 1 \ |\ j^* = j) = \eta$ for some known $\eta > 0.5$,
while the distribution of rewards of all other options is $P(r_{j,t} =
1 \ |\ j^* \neq j) = 0.5$.  These distributional assumptions represent
a setting in which most options are on average neither beneficial nor
harmful but one may be slightly better than the others, which roughly
approximates the setting of our data since most users have near zero
ROI.  As noted below, this model is straightforward to generalize to
bounded continuous-valued rewards and arbitrary $P(r_{j,t} \ |\ j^* =
j)$ and $P(r_{j,t} \ |\ j^* \neq j)$ satisfying the monotone
likelihood ratio property.

We further assume that at each point in time, each decision-maker will
make a decision to commit to one specific option.  When committing to
an option, individuals are incentivized to choose the options they
think are best because on the time step following each of their
decisions, each decision-maker will be rewarded according to the
reward signal of the option that the decision-maker chose in the
previous time step.  At the point of decision in time step $t$, the
decision-makers are able to observe the history of public signals
$r_{j,1}, \ldots, r_{j,t}$ for each option, as well as the number of
decision-makers who chose each option in the last time step, i.e. the
previous popularity of option $j$ at time $t$, which we denote as
$p_{j,t}$.  We assume that in making their decisions, individuals will
probability match on the perceived probability that an option is the
best option.  That is, we assume individuals will commit to option $j$
at time $t$ with probability equal to $P(j^* = j
\ |\ \mathbf{r}_{\cdot, \leq t})$, where $\mathbf{r}_{\cdot, \leq t}$
is the collection of reward signals of all options up to and including
time $t$.  This assumption of probability matching is grounded in
previous theory and empirical evidence: Probability matching is a
widely observed behavior in humans and other animal species, and also
enjoys its own theoretical justifications
\cite{arganda_common_2012,vul2014one}.

However, we suppose that rather than computing the full posterior
distribution for each option, decision-makers use the following
approximation:
$$
\frac{p_{j,t}}{N} \approx P(j^* = j \ |\  \mathbf{r}_{\cdot,\leq t-1}).
$$ This approximation follows from the fact that when the entire
population is probability matching on the posterior that each option
is best, the resulting popularities of each option are then expected
to be proportional to those posterior probabilities, i.e. $E[p_{j,t}]
= N \cdot P(j^* = j \ |\ \mathbf{r}_{\cdot,\leq t-1})$. Thus in this case the
posterior from the last time step can be approximately recovered from
popularity.  This approximation is reasonable if the population of
decision-makers $N$ is large compared to the number of options $M$.

Using this approximation, Bayes' rule then indicates that
decision-makers will probability match according to the probability
$Q(j^* = j \ |\ r_{\cdot,t}, p_{\cdot,t})$,
$$
Q(j^* = j \ |\  r_{\cdot,t}, p_{\cdot,t}) =
\frac{
  P(r_{j,t} \ |\  j^* = j) \cdot p_{j,t}
}{
  \sum_k P(r_{k,t} \ |\  j^* = k) \cdot p_{k,t}
} \approx P(j^* = j\ |\  \mathbf{r}_{\cdot, \leq t}),
$$ where the simplification in the likelihood term comes from the
assumed form of the reward distributions.  Therefore, rather than
spending valuable computation time to incorporate the entire history
of reward signals, decision-makers use social information to
approximate this distribution by incorporating popularity as a prior
distribution on the probability that an option is best.

Decision-makers can sample from this probability distribution
efficiently by using the following strategy.  The decision-maker first
samples an option to consider according to popularity. The
decision-maker then commits to that option with probability
proportional to the likelihood that the option is best given just its
latest reward signal.  If the decision-maker repeats this procedure
until an option is chosen, the resulting decision probabilities will
be as desired.

As a final note, in order to use the above specification of the social
sampling model in our analysis of the eToro data, we must discretize
the continuous performance signals obtained from the eToro data.
Given the performance signal $q_{j,t}$ of user $j$ on day $t$ from the
eToro data, we simply let $r_{j,t} = \textbf{1}(q_{j,t} > 0)$, where
$\textbf{1}$ is an indicator function.  The model above can then be
used on these coarsened signals.  Beyond this need to discretize, the
main differences between the model specification above and the actual
environment on eToro are that the performance signals on eToro will be
correlated from day-to-day and that on eToro the number agents $N$ and
the number of options $M$ change over time.  Nevertheless, we can use
the model without modification in this setting.

\paragraph{Note: Model Extensions}
While the social sampling model derived above is the version we use
for the purposes of our work, this model makes a number of simplifying
assumptions.  The social sampling model can easily be extended to
relax some of these assumptions, and deeper exploration of these
extensions merits future work.

The social sampling model only specifies how users gain mimickers.
Another important characteristic of our data that this model ignores
is the fact that once a user makes a decision to mimic another user,
that decision must be explicitly reverted.
However, it is straightforward to extend the social sampling model in
this direction. To do this, we can add a component of the social
sampling model to allow decision-makers to revoke previous decisions.
In this setting, there are still $N$ new decision-makers on each day
making decisions according to the previously specified model, but we
also suppose that all previous decisions to commit to option $j$ are
maintained on each day with probability equal to (or proportional to)
$P(r_{j,t} \ |\ j^* = j)$.  We then have that the expected total
popularity on each day will be given by
$$
E[p_{j,t+1}] =  p_{j,t} \cdot P(r_{j,t} \ |\ j^* = j) + N \cdot Q(j^* = j \ |\  r_{\cdot,t}, p_{\cdot,t}),
$$ which is the sum of the remaining decisions to take that option
from the previous day plus the new decisions to take that option.
Substituting and simplifying this becomes,
$$
E[p_{j,t+1}] =
\left(
N + \sum_k P( j^* = k \ |\ r_{k,t}) \cdot p_{k,t}
\right)
Q(j^* = j \ |\  r_{\cdot,t}, p_{\cdot,t})
$$
$$
\propto
Q(j^* = j \ |\  r_{\cdot,t}, p_{\cdot,t})
 \approx P(j^* = j\ |\  \mathbf{r}_{\cdot, \leq t}),
$$ if as above $p_{j,t}/\sum_k p_{k,t} \approx P(j = j^*
 \ |\ \mathbf{r}_{\cdot, < t})$.  Thus, popularity will be expected to
 continue to approximate the posterior each option is best if
 decision-makers maintain their previous decisions according to recent
 signals of quality.  This mechanism is highly plausible in the case of
 eToro.  Users may initially decide whom to mimic by considering
 options according to popularity and committing according to recent
 signals of quality, but users may decide to continue mimicking simply by
 paying attention to whether they lose or gain money in a mimic
 relationship.  However, preliminary analyses were inconclusive as to
 whether our dataset supports this model of ``unfollowing'' behavior.

The social sampling model and its justification are also
straightforward to generalize to more complicated reward
distributions.  For arbitrary $P(r_{j,t} \ |\ j^* = j)$ and $P(r_{j,t}
\ |\ j^* \neq j)$ with real-valued $r_{j,t}$ we will instead have
$$
Q(j^* = j \ |\  r_{\cdot,t}, p_{\cdot,t})
\propto
\frac{
  P(j^* = j  \ |\ r_{j,t})
}{
  P(j^* \neq j  \ |\ r_{j,t})
} \cdot p_{j,t}
\propto
\frac{
  P(r_{j,t}  \ |\ j^* = j)
}{
  P(r_{j,t}  \ |\ j^* \neq j)
} \cdot p_{j,t},
$$
still approximating $P(j^* = j\ |\  \mathbf{r}_{\cdot, \leq t})$.
Assuming $r_{j,t}$ is upper bounded by $R$, and assuming $P(r_{j,t}
\ |\ j^* = j)$ and $P(r_{j,t} \ |\ j^* \neq j)$ satisfy the monotone
likelihood ratio property, then
$$
\frac{
  P(r_{j,t}  \ |\ j^* = j)
}{
  P(r_{j,t}  \ |\ j^* \neq j)
}
\leq
\frac{
  P(r_{j,t} = R  \ |\ j^* = j)
}{
  P(r_{j,t} = R \ |\ j^* \neq j)
} = C,
$$ where $C$ is hence a constant upper bound on the likelihood ratio.
Probability matching on this quantity can be achieved by first
sampling according to popularity and then committing to that decision
with probability
$$
\frac{1}{C} \frac{
  P(r_{j,t}  \ |\ j^* = j)
}{
  P(r_{j,t}  \ |\ j^* \neq j)
}.
$$

\subsection*{Model Comparison}

To further assess how well the social sampling model fits our data we
compare the errors in the predictions of this model to those of the
alternative models.  We first compare average prediction error using
multiple error metrics and three methods of cross-validation.
Cross-validation consists of holding out subsets of data, fitting all
of the models to the non-held out data, and then testing the
fine-grained predictions those models make on the held out data.  We
use three different forms of cross-validation that differ in the ways
held out sets are determined.  First, we use a 10-fold
cross-validation where we form a random partition of all of the users
in our data into 10 groups.  We then use 9 out of 10 groups for
training and the remaining group for testing, repeated 10 times with
each group of users held out once.  Next, we use a similar 10-fold
cross-validation method where we instead form a random partition of
all the days in our data into 10 groups.  Finally, rather than holding
out random days, we hold out the most recent set of days containing
approximately 10\% of our data, training on the earlier 90\%.  Using
these particular forms of cross-validation helps to ensure that any
differences in predictive performance are not just due to there being
only a few users or a few days that we can model well.

We use four different error metrics to quantify prediction error: mean
absolute error (MAE), mean squared error (MSE), and F-score.  Mean
squared error and mean absolute error are defined as the sum of the
squared values of the prediction residuals and the sum of the absolute
values of the prediction residuals, respectively.  (The prediction
residuals are given by the difference between the predicted numbers of
new mimickers minus the actual number for each user on each day.)
F-score quantifies the error in predicting binary response variables.
In our case, we attempt to predict whether a particular user has
greater than zero new mimickers as the response variable for these
metrics.  For the F-score, we round each model's predictions to the
nearest whole number and classify a user as having greater than zero
mimickers if the predicted expected number of mimickers according to the
model is greater than 0.

The results of these three methods of cross-validation for each error
metric are presented in Tables S3-S5.
The social sampling model is dominant in terms of mean absolute error
and F-score.  However, the additive model has the best mean-squared
error in two of three cases.  The fact that social sampling achieves
good mean absolute error but sometimes worse mean squared error
indicates that the inferior performance of social sampling on mean
squared error may be driven by a small fraction of data points.  This
interpretation is also supported by an additional set of results
presented in Figure S1.  These results indicate that the social
sampling model achieves lower error on a majority of data points
(using the most recent 10\% hold-out method).

\begin{figure*}
  \centering
  \includegraphics[width = 0.45\linewidth]{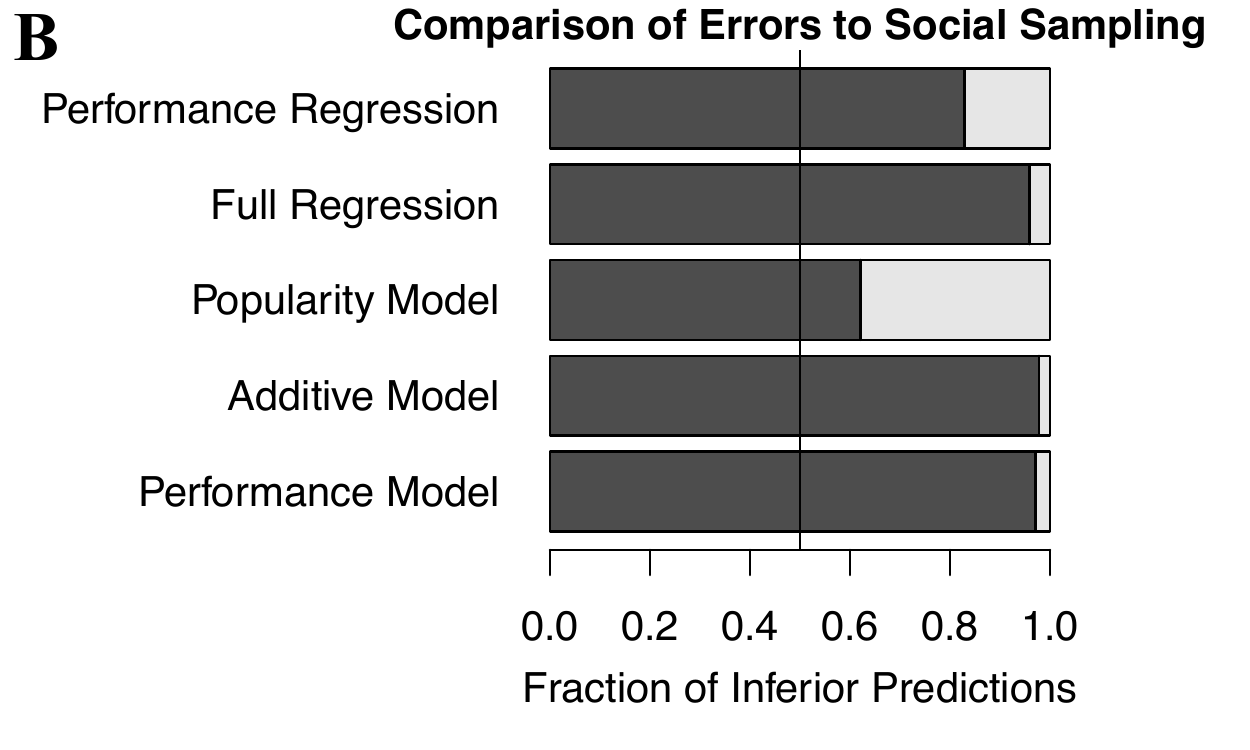}
  \caption{The horizontal bar plot indicates the relative error of
    each alternative model as compared to the social sampling model.
    Relative error is measured as the fraction of the alternative
    models' prediction residuals that are larger than those of the
    social sampling model.}
\end{figure*}

Finally, to better understand the relationship between these
quantitative model comparison results and the qualitative results from
Figure 2 in the main text, we further explored the distribution of
model prediction errors for the three best models: social sampling,
the popularity model, and the additive model.  In this case we look at
in-sample error after training on the entire dataset, which produces
similar results to the cross-validation methods since the amount of
data we have is so large compared to the number of parameters of each
model.  We examined the error of each model on the ten subsets of data
defined by the data binning scheme used to generate the plots in
Figure 2 of the main text.  We find that for many subsets of the data
(generally for users with positive performance, and for users with 10
mimickers or fewer), the predictions of these three models tend to be
quite similar, and these three models achieve MSE and MAE within about
5\% or less of each other on each of these subsets.  However, for
users with more than 100 mimickers and non-positive performance,
social sampling achieves about 35\% better MSE and MAE.  Similarly,
for users with between 10 and 100 mimickers and non-positive
performance, social sampling achieves about 35\% better MAE and about
10\% better MSE.  Since the dataset is unbalanced, the large gains in
predictive performance that the social sampling model has on these
subsets of the data are likely washed out by the many data points with
low popularity or high popularity and high performance.  This
interpretation is consistent with the patterns evident in Figure 2 of
the main text.

\begin{table}[h]
  \centering
  \caption{Results using mean absolute error. Lower mean absolute error values are better.}
  \begin{tabular}{|l|c|c|c|}
    \hline
    \textbf{Model} & \textbf{By User} & \textbf{By Day} & \textbf{In Future} \\ \hline
    Social Sampling & \textbf{0.0391} & \textbf{0.0455} & \textbf{0.0403} \\ \hline
    Performance Model & 0.0782 & 0.0783 & 0.0626 \\ \hline
    Additive Model & 0.0402 & 0.0478 & 0.0420 \\ \hline
    Popularity Model & 0.0400 & 0.0465 & 0.0409 \\ \hline
    Full Regression & 0.0527 & 0.0569 & 0.0480 \\ \hline
    Performance Regression & 0.0782 & 0.0783 & 0.0625 \\ \hline
  \end{tabular}
\end{table}

\begin{table}[h]
  \centering
  \caption{Results using mean squared error. Lower mean squared error values are better.}
  \begin{tabular}{|l|c|c|c|}
    \hline
    \textbf{Model} & \textbf{By User} & \textbf{By Day} & \textbf{In Future} \\ \hline
    Social Sampling & \textbf{1.2924} & 2.1673 & 2.2389 \\ \hline
    Performance Model & 3.0727 & 3.0683 & 2.3864 \\ \hline
    Additive Model & 1.3194 & \textbf{2.1294} & \textbf{2.0707} \\ \hline
    Popularity Model & 1.3215 & 2.1768 & 2.1428 \\ \hline
    Full Regression & 2.1291 & 2.6120 & 2.0933 \\ \hline
    Performance Regression & 3.0728 & 3.0682 & 2.3858 \\ \hline
  \end{tabular}
\end{table}

\begin{table}[h]
  \centering
  \caption{Results using F-score. Higher F-score values are better.}
  \begin{tabular}{|l|c|c|c|}
    \hline
    \textbf{Model} & \textbf{By User} & \textbf{By Day} & \textbf{In Future} \\ \hline
    Social Sampling & \textbf{0.3978} & \textbf{0.4061} & \textbf{0.3619} \\ \hline
    Performance Model & 0.0073 & 0.0000 & 0.0000 \\ \hline
    Additive Model & 0.3874 & 0.3944 & 0.3418 \\ \hline
    Popularity Model & 0.3873 & 0.3941 & 0.3446 \\ \hline
    Full Regression & 0.3751 & 0.3775 & 0.3537 \\ \hline
    Performance Regression & 0.0123 & 0.0007 & 0.0000 \\ \hline
  \end{tabular}
\end{table}

\subsection*{Robustness Checks}

Finally, to further test the robustness of all our results to our
choices in data parsing, we run the same set of analyses described in
the previous sections using several alternative parses of our data.
We test robustness to (a) the size of the rolling window used in our
expected ROI computations, (b) the type of performance metric we use,
(c) whether we use our imputed data or not, (d) whether we include
only closed trades in our performance computation, and (e) which
columns in our data we rely on for determining amounts invested and
profit.  We also check the results in our held-out year of data.

To test robustness to the performance metric used, we consider several
alternatives.  With our main ROI performance metric we check three
alternative rolling window sizes of 1 day, 60 days, and 90 days.  We
also use two risk-adjusted performance metrics based on the Sharpe and
Sortino ratios.  The Sharpe metric is simply the expected ROI from
closed trades divided by the standard deviation in those returns.  The
Sortino metric is the expected ROI divided by the standard deviation
in the negative returns.
Additionally, we consider the sum of profit from all trades (the ``sum
metric''), the sum of the profit from all trades divided by the total
amount invested in all trades (the ``average metric''), and the
percentage of trades with positive profit minus 0.5 (the ``percent
metric'').  For all these alternative metrics we use 30-day rolling
windows.

For the other robustness checks, we maintain our usual 30-day ROI
metric.  To check robustness to whether we impute trades, we check the
results without using imputed trades.  To test robustness to whether
we only use closed trades, we compute average ROI after simulating a
liquidation of all open trades at the end of each day (in the same way
as described in the section ``Data Processing'' above).  To test
robustness to which columns of our data we rely on, we recompute the
profit from each trade and the amount invested in each trade via the
total units purchased in those trades.  We also use one additional ROI
performance metric using the profit according to the units divided
directly by the units invested to account for leverage.

Here we focus on an aggregate analysis of the results from these
alternative data parses.  We highlight the places where we have
identified deviations from the results of our main analyses.  In 12
out of 13 of these alternative parses we observe a statistically
significant positive multiplicative interaction between previous
popularity and previous performance in relation to changes in
popularity using the ordinary linear model.  The one exception is the
parse using the Sortino metric.

The model comparison results are also highly robust, though slightly
more mixed.  In one case out of 39 total cases (counting the user,
day, and future cross-validation approaches separately for each of the
13 different parses), the social sampling model has inferior mean
absolute error.  In four of the 12 alternative parses (the parses
using rolling window 60, no data imputation, liquidating open trades,
and the average metric), social sampling is inferior on mean squared
error in the single method of cross validation that it was superior on
in our main analysis.  In two out of the 39 cases, social sampling has
worse F-score (holding out the latest 10\% using rolling window of 7
and using the percent metric).  The relative error results are
qualitatively similar to our main results for all models and all
parses except that the popularity model achieves superior relative
error in three out of the thirteen cases (rolling window 1, rolling
window 7, and the percent metric).

The tests of the specific parameter fits of the social sampling model
were also quite robust across many alternative parses.  In every
alternative parse, a scaling factor of 1.0 is the best fit exponential
scaling factor on popularity, indicating strong evidence that
popularity is used linearly rather than sublinearly or superlinearly.
In every alternative parse at least one of the top ten traders'
confidence interval's contains the best skill parameter according to
the model.  In four cases (percent metric, and rolling windows 1, 60,
and 90), only three of the confidence intervals for the top 10
traders' skills include the inferred best skill value.  In one case
(Sortino metric) six do.  In the other eight alternative parses,
between 7 and 10 do.

%% Our prediction results comparing the returns of the group portfolios
%% according to the social sampling and the performance models are also
%% robust.  In only one parse out of 13 (90-day rolling window ROI) is
%% the mean of the distribution of the daily return from the group
%% portfolio lower using the social sampling model, and in all cases the
%% standard deviation is higher.  In all cases when the mean is higher,
%% it is between roughly 15\% and 50\% higher, typically close to 25\%.
%% The standard deviation is always between roughly 75\% and 130\%
%% higher.  The Mann-Whitney $U$ tests are all significant at the 0.01
%% level (and in most cases at the 0.001 level).

The results on our held-out validation set were similar to the
aggregate results from the alternative parses, though somewhat weaker.
As described in Section ``Data Processing'' above, we reserved one
year of data as a nearly completely held-out validation set.  This
year ranged from July 1, 2012 until June 30, 2013.  On this held out
validation set, we used the 30-day ROI performance metric as in our
main analysis.  First, we again found evidence for the positive
multiplicative interaction effect using the ordinary linear model.
Next, the model comparison results were similar to those in our main
analysis but somewhat weaker.  The social sampling model again
dominated in terms of mean absolute error, but lost in one out of
three cases of the F-score and lost in all cases in mean squared
error.  Further, the relative error results were also similar to but
somewhat weaker than the results in our main analysis.  Once again
social sampling achieves better error on more that 90\% of cases
compared with every model except the popularity model. The popularity
model actually achieves better error on about 68\% of the predictions,
therefore beating social sampling under this relative error metric on
the held out set.  When performing a more fine-grained analysis of the
errors, as we did with our model comparison results from the first
year of data, we find once again that the predictions of the social
sampling model, the popularity model, and the additive model are
similar for many subsets of the data.  However, the social sampling
model predicts reliably better on low performance traders and
substantially better on high popularity, low performance traders.
Preferential attachment and the additive model do have noticeably
better predictions on the single subset of traders with very high
popularity and high performance, but still these gains are small
compared to those of social sampling on the subsets where social
sampling predicts well.  As such, while the model comparison results
on the held out set are slightly weaker, once again it appears that
the highly unbalanced data makes the average errors overall look much
different from the average errors within these relevant subsets of the
data.
The tests of the specific fit of the parameters of the social sampling
model were highly robust in the held out set.  Once again, the best
fit scaling exponent on popularity was a linear fit.  All ten of the
top ten traders' confidence interval's included the inferred best
skill value according to the model.
%% Finally, the predicted group
%% performance results comparing social sampling to the performance model
%% did not replicate in the held-out set.  We find that the standard
%% deviation of the return on the group portfolio increases by about 40\%
%% in the social sampling model versus the performance model.  However,
%% the expected return is about 5\% lower under social sampling that
%% under the performance model rather than being higher, and the
%% Mann-Whitney $U$ test of the difference between the two distributions
%% is not significant.
Overall, our results are highly robust across
alternative parses of the data, and moderately robust on the held-out
data.

\subsection*{Possible Confounding Factors}

\begin{figure*}
  \centering \includegraphics[width = 0.6\linewidth]{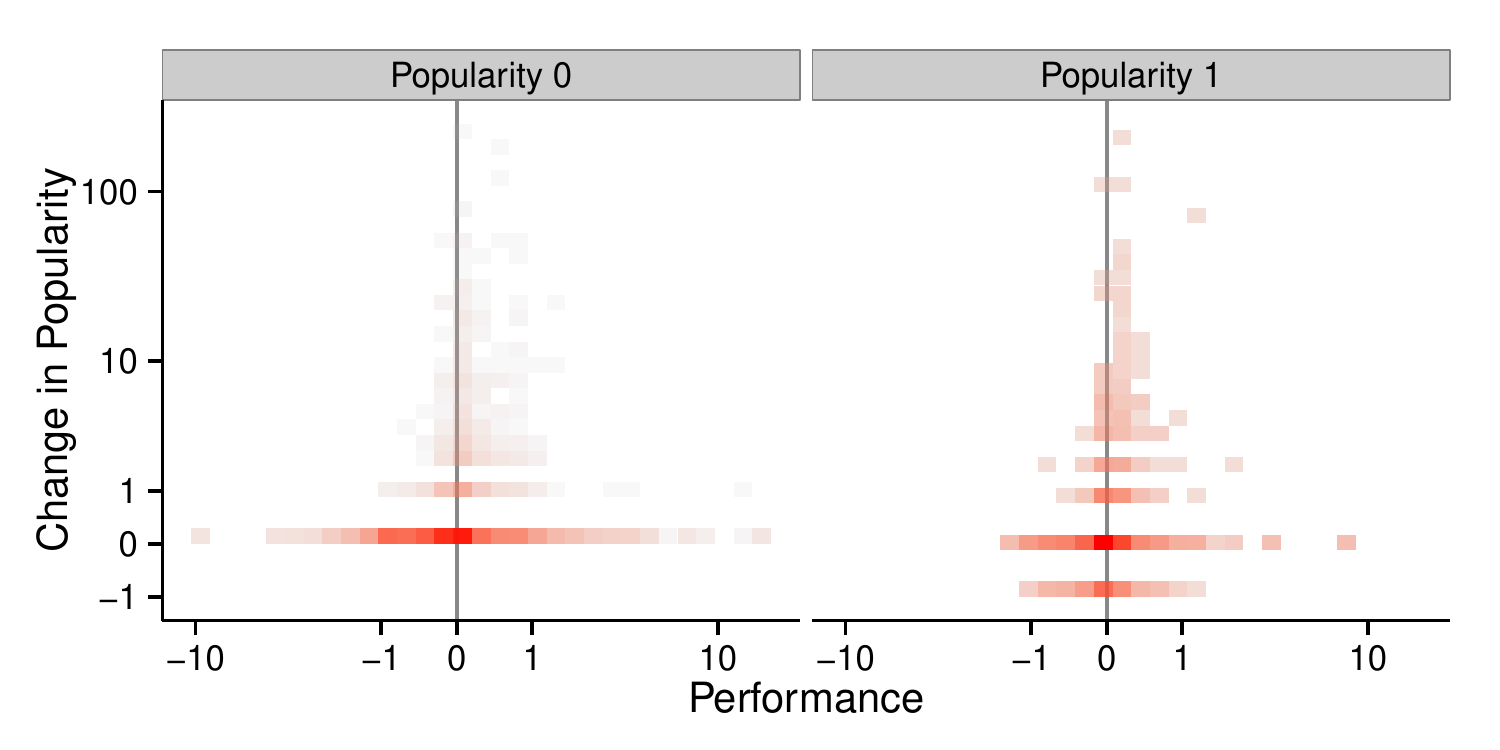}
  \caption{Evidence for a multiplicative interaction between
    popularity and performance in relation to changes in popularity
    even at low levels of popularity. The left panel includes all
    users on all days they have 0 mimickers, while the right panel
    includes all users on all days they had exactly 1 mimicker.  The
    plots themselves are two-dimensional histograms (binned scatter
    plots).  Higher red-value and lower transparency indicate higher
    density.  The plot shows that users with one mimicker are expected
    to have greater increases in popularity than users with zero
    mimickers, and that this difference is mainly present in individuals
    with non-negative performance.  As in the main text, the
    performance metric here is a 30-day rolling average of daily ROI,
    and change in popularity is the change in popularity on the
    following day.  Running the regressions from the ``Data Analysis''
    section above on the subset of data consisting of users with
    exactly zero or one mimicker yields a significant positive
    interaction term ($p < 0.05$) with either ordinary or robust
    standard errors.}
\end{figure*}

Our statistical analysis and model comparison allowed us to provide
evidence that our results are likely not due to our apparent observed
effects holding for just a few users, or on just a few days.  Another
possible limitation of our analysis is that we attempt to infer a
causal mechanism from observational data.  It is clear that changes in
popularity cannot be causing previous popularity or previous
performance, so a non-causal interpretation of our results would have
to involve an unobserved variable.  Such an unobserved variable would
have to be correlated with both popularity and performance to cause
the multiplicative interaction we observe.  Among the possible
alternative explanations of our results, the most likely is position
bias.  The issue of confounding position bias with social influence
has been problematic for many previous related studies.  A position
bias occurs when a decision-maker is presented with a list of options.
Those items at the top of the list tend to receive more attention and
be selected more often simply because they are at the top of the list,
not because they are any better in terms of underlying quality
\cite{hodas_simple_2014}.  Thus, if the list is ordered by current
popularity, a position bias can make it appear that highly popular
items are being chosen more often.

There are two reasons that a position bias is unlikely to be
responsible for our results.  First and foremost, while it is possible
for a position bias to account for a marginal effect of previous
popularity on future popularity, position bias resulting from a single
ordered list would not cause the multiplicative interaction we
observe.  Users would have to either have sorted by performance and
popularity simultaneously for the position bias to cause that
interaction, or some other feature of the interface would have to
highlight users who were both popular and high performing.  To the
best of our knowledge, there was no straightforward way to sort by
these two features simultaneously in the eToro interface at the time
our data was collected.  Furthermore, as shown in Figure S2, we
observe the interaction between popularity and performance even at
very low levels of popularity.  While some users undoubtedly
persistently scroll through the list of most popular traders down to
those who have just one mimicker, a likely more common occurrence that
could explain this interaction is that users sort according to a
performance metric then prefer to consider mimicking users that have
one mimicker as opposed to zero.

\subsection*{Anecdotal Evidence}

One final piece of evidence supporting the social sampling model is a
set of self-reports of actual users of eToro.  Several users have
posted advice online about how to effectively find traders to mimic on
eToro. Two personal accounts we examined in detail\footnote{See ``How
  to Choose Good Traders or Gurus to Copy in Etoro? - Video Guide''
  from YouTube (https://www.youtube.com/watch?v=di7Sw587has, 3:02),
  and ``eToro Tips: Find Best Gurus'' from SocialTradingGuru.com
  (http://socialtradingguru.com/tips/etoro-tips/select-etoro-gurus).}
broadly match the model we propose.  The online advice points to
sorting traders by popularity as the most reliable way to find
potential traders to mimic, and the advice emphasizes that you should
look carefully at a trader's profile before deciding to mimic that
trader to better judge the trader's performance, even if the trader is
highly popular.  Both accounts also note that it's not just the top
most mimicked traders who are potentially good to mimic but that it is
also productive to look for less popular options.

\end{document}